\documentclass[aps,prd,superscriptaddress,showpacs,preprint,amsmath,amssymb]{revtex4}
\usepackage{graphicx, bm}
\usepackage{float}
\usepackage[usenames]{color}

\begin{document}

\draft

\title{Study of the projected sensitivity on the anomalous quartic gauge couplings via $Z\gamma\gamma$ production at the CLIC }

\author{V. Ari\footnote{vari@science.ankara.edu.tr}}
\affiliation{\small Department of Physics, Ankara University, Turkey.\\}

\author{E. Gurkanli\footnote{egurkanli@sinop.edu.tr}}
\affiliation{\small Department of Physics, Sinop University, Turkey.\\}

\author{M. K\"{o}ksal\footnote{mkoksal@cumhuriyet.edu.tr}}
\affiliation{\small Department of Physics, Sivas Cumhuriyet University, 58140, Sivas, Turkey.}

\author{ A. Guti\'errez-Rodr\'{\i}guez\footnote{alexgu@fisica.uaz.edu.mx}}
\affiliation{\small Facultad de F\'{\i}sica, Universidad Aut\'onoma de Zacatecas\\
         Apartado Postal C-580, 98060 Zacatecas, M\'exico.\\}

\author{ M. A. Hern\'andez-Ru\'{\i}z\footnote{mahernan@uaz.edu.mx}}
\affiliation{\small Unidad Acad\'emica de Ciencias Qu\'{\i}micas, Universidad Aut\'onoma de Zacatecas\\
         Apartado Postal C-585, 98060 Zacatecas, M\'exico.\\}

\date{\today}

\begin{abstract}

Gauge boson self-couplings are completely defined by the non-Abelian gauge symmetry of the Standard Model (SM). For this reason,
a direct search for these couplings is extremely significant in understanding the gauge structure of the SM. The possible deviation
from the SM predictions of gauge boson self-couplings would be a sign of the presence of new physics beyond the SM. In this work,
we study the sensitivities on the anomalous couplings defined by dimension-8 operators related to the $ZZ\gamma\gamma$ and
$Z\gamma\gamma\gamma$ quartic vertices through the process $e^+e^- \to Z\gamma\gamma$ with Z-boson decaying to charged leptons at the Compact Linear Collider (CLIC). We analyze the center-of-mass energy of 3 TeV, integrated luminosities of ${\cal L}=1, 4, 5$ $\rm ab^{-1}$, systematic uncertainties of $\delta_{sys}=0\%, 3\%, 5\%$, unpolarized and polarized electron beams and considering the Initial State Radiation and Beamstrahlung effects for extraction of expected sensitivity on the anomalous $f_ {T,j}/\Lambda^4$ couplings at $95\%$ confidence level, which are especially sensitive to the $Z\gamma\gamma$ channel. It is clear from the results that we expect better limits on the couplings if the systematic error is improved. The best limits obtained on the anomalous quartic couplings for the process $e^+e^- \to Z\gamma\gamma$ with $\sqrt{s}= 3$ TeV, ${\cal L}=5$ $\rm ab^{-1}$, and $\delta_{sys}=0\%$ can be approximately improved up to about 1.1 times better than the limits obtained with $\delta_{sys}=5\%$. Our sensitivities on the anomalous quartic couplings can set more stringent sensitivity by two orders of magnitude concerning the best sensitivity derived from the current experimental limits. Finally, with initial electron beam polarization, the sensitivity of the anomalous quartic couplings improves by almost a factor of 1.2.

\end{abstract}

\pacs{12.60.-i, 14.70.Hp, 14.70.Bh \\
Keywords: Electroweak interaction, Models beyond the Standard Model, Anomalous couplings.\\
}

\vspace{5mm}

\maketitle


\section{Introduction}

In the Standard Model (SM) of elementary particle physics, the structure of the fermion-vector boson couplings has been extensively
tested \cite{Glashow,Weinberg,Salam,Data2020}. However, other sectors of the SM, such as the one corresponding to the self-couplings
of the vector bosons, such as triple $VVV$ and quartic $VVVV$ couplings with $V= \gamma, Z, W^{\pm}$ have been less tested.
In addition, the study of the neutral gauge boson self-couplings provides one of the most sensitive probes of new physics
beyond the SM (BSM) \cite{Data2020}.

In the SM, the non-Abelian nature of the $SU(2)_L \times U(1)_Y$ gauge group of the electroweak interactions leads
to gauge boson self-interactions through the Triple Gauge Couplings (TGC) and the Quartic Gauge Couplings (QGC).
While the anomalous TGC (aTGC) and the anomalous QGC (aQGC) are deviations from the SM. Whilst both aTGC and aQGC
offer an important test of the non-Abelian gauge structure of the SM, the aQGCs are also connected to the electroweak
symmetry breaking sector. In addition, the aQGC can be regarded as a window on the mechanism responsible for the
electroweak symmetry-breaking \cite{Data2022}. The SM predicts only QGCs that contain at least two charged $W$ bosons
$(W^+W^-W^+W^-, W^+W^-ZZ,W^+W^-\gamma\gamma,W^+W^-Z\gamma)$, while the neutral gauge boson couplings $(ZZZZ, ZZZ\gamma,
ZZ\gamma\gamma, Z\gamma\gamma\gamma, \gamma\gamma\gamma\gamma)$ are excluded and, if exist, they are considered to be
effects of physics BSM. In the particular case of the production of $Z\gamma\gamma$ tribosons, a suitable and relatively
modern approach to observe the effects of new physics BSM is to use an effective Lagrangian description of the SM. Physics
beyond the SM could induce these anomalous neutral QGCs, enhancing the cross section for $Z$ production and modifying the
kinematic distribution of the final-state $Z$ boson and photons. The effect of such new couplings can be modeled using
an effective field theory (EFT) that includes higher-dimensional operators. In a recent paper \cite{CPC44-2020} it is discussed 
how the neutral Triple Gauge Couplings (nTGC) $ZV\gamma$ $(V=\gamma,Z)$ can be generated by effective dimension-8 operators, 
and then present cross-sections for $e^+e^- \to Z\gamma$ production. In another paper \cite{PRD107-2023}, the authors study 
probes of nTGC in the reactions $pp(q\bar q) \to Z\gamma$ at the LHC and the proposed 100 TeV $pp$ colliders, and compare 
their sensitivity reaches with those of the proposed $e^{+}e^{-}$ colliders. See Refs. \cite{Eboli-PRD93-2016,Marantis-JPCS2105-2021,Degrande-AP335-2013,Hao-PRD104-2021,Degrande-JHEP02-2014,Hays-JHEP02-2019,
Gounaris-PRD65-2002,Wrishik-JHEP08-2022,Murphy-JHEP10-2020,Ellis-China64-2021,Murphy-JHEP04-2021,Green-RMP89-2017}
for a summary of the study of the aQGC in the context of dimension-8 gauge-invariant operators.

The physics program of the ATLAS and CMS Collaborations at the CERN Large Hadron Collider (LHC) contemplates the study of the physics
of the anomalous Neutral Quartic Gauge Couplings (aNQGC) and the anomalous Neutral Triple Gauge Couplings (aNTGC), which are deviations 
from the SM \cite{PRD93-2016,JHEP10-2017,JHEP10-2021,arXiv:2106.11082}. Actually both ATLAS \cite{PLB760-2016} and CMS \cite{JHEP12-2018} 
Collaborations already measured NTGC. In the following, some studies on future results of the ATLAS
experiment on aQGC are presented. The L3 and OPAL Collaborations at the CERN LEP-II reported for the first time bounds
on the aQGC $ZZ\gamma\gamma$ coupling via $e^+e^-$ collisions \cite{PLB540-2002,PRD70-2004}. Other studies are realized
in hadron-hadron colliders by the D0 and CDF Collaborations at the Tevatron at Fermilab \cite{PRD62-2000,PRD88-2013}
and by the ATLAS \cite{PRL113-2014,PRL115-2015,PRD96-2017,EPJC77-2017,PRD95-2017} and CMS Collaborations \cite{PRD90-2014,PLB774-2017,PLB770-2017,JHEP06-2017,PRL120-2018,PLB795-2019,PLB798-2019,PLB812-2021,JHEP06-2020,PLB809-2020,PLB811-2020}
at the LHC. Other processes and potential mechanisms for the phenomenological analyses of anomalous quartic neutral interactions
are the followings, $e^+e^- \to Z\gamma\gamma$ \cite{Stirling,PRD89-2014}, $e^+e^- \to ZZ\gamma$ \cite{EPJC13-2000},
$e^+e^- \to qq\gamma\gamma$ \cite{PLB515-2001}, $e^-\gamma \to ZZe^-$ \cite{Eboli}, $e^-\gamma \to Z\gamma e^-$ \cite{PRD75-2007},
$\gamma\gamma \to ZZ$ \cite{AHEP2016-2016}, $\gamma\gamma \to \gamma Z$ \cite{JHEP10-121-2021}, $\gamma\gamma \to \gamma \gamma$ \cite{EPJC81-2021},
$e^+e^- \to ZZ\gamma$ \cite{EPJP130-2015}, $pp \to Z\gamma jj$ \cite{PRD104-2021}, $pp \to p\gamma\gamma p \to pZ\gamma p$ \cite{JHEP06-142-2017},
$p\bar p \to Z\gamma\gamma$ \cite{JPG26-2000}, $pp(\bar p) \to \gamma\gamma ll$ \cite{Eboli2}, $pp \to p\gamma^*\gamma^* p \to pZZ p$ \cite{PRD81-2010,PRD85-2012,arXiv:0908.2020}, $pp \to p\gamma^* p \to P ZZqX$ \cite{arXiv:1311.1370}, $pp \to p\gamma^*p \to p\gamma ZqX$
\cite{PRD86-2012}, $pp \to qq\gamma ll$ \cite{Eboli4}, $pp \to Z\gamma\gamma$ \cite{arXiv:2109.12572}. For previous work on aQGC at linear
colliders, we refer the reader to see Refs. \cite{Data2020,Han-PLB1998,Daniele-RNC1997, Boos-PRD1998,Boos-PRD2000,Beyer-EPJC2006,Christian-EPJC2017,Stephen-arxiv:9505252,
Stirling-PLB1999,Belanger-PLB1992,Senol-AHEP2017,Barger-PRD1995,Cuypers-PLB1995,Cuypers-IJMPA1996,Ji-arXiv:2204.08195} for an overview
of the phenomenological status.

It is appropriate to mention that the ATLAS and CMS Collaborations recently observed the $pp \to Z\gamma\gamma \to l^+l^-\gamma\gamma$
production channel with a center-of-mass energy of 8 TeV and integrated luminosity of 20.3 ${\rm fb}^{-1}$, respectively \cite{PRD93-2016,JHEP10-2017}.
More recently, the first measurements of the $pp \to Z\gamma\gamma \to l^+l^-\gamma\gamma$ cross-section at $\sqrt{s} = 13$ TeV
is reported by the CMS Collaboration, corresponding to an integrated luminosity of ${\cal L}=137$ $\rm fb^{-1}$ \cite{JHEP10-2021}. In that reference,
the best sensitivities of the aQGC $f_{T,j}/\Lambda^4$ ($j=0,1,2,5,6,7,8,9$) are $[-5.70;5.46]$, $[-5.70;5.46]$, $[-11.4;10.9]$, $[-2.92;2.92]$, $[-3.80;3.88]$, $[-7.88;7.72]$, $[-1.06;1.10]$ and $[-1.82;1.82]$ TeV$^{-4}$, respectively.

Due to the nature of the colliding particles, lepton colliders are generally helpful for examining the processes. Because the elementary particles are fully defined at the fundamental level, the collisions in the lepton collider are clean without any hadronic activity, and the measurements are made more precisely. The hadron colliders have higher collision energies than the lepton colliders. However, high collision energies are crucial in investigating the new particles and their interactions. On the other hand, each collision in a hadron collider composes several backgrounds for physics analysis, creating a large number of elementary processes. Therefore it is not easy to analyze and make precise measurements in hadron colliders. The discovery potential of the LHC would be complemented by the CLIC, which is a  multi-TeV high-luminosity linear $e^+e^-$ collider under development.

The current CLIC baseline staging scenario assumes 1.0 $\rm ab^{-1}$, 2.5 $\rm ab^{-1}$, and 5.0 $\rm ab^{-1}$ of luminosity at the three energy stages of $\sqrt{s}=0.38$ TeV, 1.5 TeV, and 3 TeV, respectively \cite{franc}. Also, the baseline CLIC provides ±80\% longitudinal polarisation for the electron beam and no positron polarisation. The expected integrated luminosities for the CLIC with the polarized electron beams are ${\cal L}=1$ ab$^{-1}$ ($P_{e^-}=80\%$), ${\cal L}=4$ ab$^{-1}$ ($P_{e^-}=-80\%$), and ${\cal L}=5$ ab$^{-1}$ ($P_{e^-}=0\%$) \cite{franc,CLIC-1812.06018}. Thanks to the clean event environment,
the tunable collision energy, and the potential to polarize beams, it is possible for the CLIC that observe the deviation from SM predictions indicating new physics, discover new particles, and make precise measurements. Furthermore, beam polarisations combined with high luminosity increase the analysis capability, resulting in better statistics, reducing systematic errors, enhancing the signal or suppressing the SM processes, and revealing new processes. Thus, using
polarized electron beams is beneficial in increasing signal rates and minimizing unwanted background processes.

The cross-section of any process is defined from the four possible pure chiral cross-sections with electron beam polarizations $P_{e^-}$ and positron beam polarization $P_{e^+}$ as follows \cite{kek2017}

\begin{eqnarray}
\label{}
\sigma\left(P_{e^-},P_{e^+}\right)=\frac{1}{4}\{\left(1+P_{e^-}\right)\left(1+P_{e^+}\right)\sigma_{RR}+\left(1-P_{e^-}\right)\left(1-P_{e^+}\right)\sigma_{LL}
\\ \nonumber
+\left(1+P_{e^-}\right)\left(1-P_{e^+}\right)\sigma_{RL}+\left(1-P_{e^-}\right)\left(1+P_{e^+}\right)\sigma_{LR}\}\,,
\end{eqnarray}

\noindent where $\sigma_{RR}$ and $\sigma_{LL}$ represent the cross-sections of both right-handed electron and positron beams and both left-handed electron and positron beams, respectively. On the other hand, $\sigma_{LR}$ and $\sigma_{RL}$ show the cross-sections of left-handed electron and right-handed positron beams and vice versa. The unpolarized cross-section $\sigma_0$ can be given as

\begin{equation}
\label{}
\sigma_0=\frac{1}{4}\{\sigma_{RR}+\sigma_{LL}+\sigma_{RL}+\sigma_{LR}\}\,.
\end{equation}

In this paper, the process $e^+e^- \to Z\gamma\gamma \to l^+l^- \gamma\gamma$ at the CLIC with unpolarized and polarized electron beams is used to investigate the presence of new physics parametrized as the aQGC between the $Z$ boson and the photons. Here, the systematic uncertainties are included in the analysis. In the final state, electron and muon channels are combined to get better sensitivity to the aQGC parameters. The paper is organized as follows. Section II
briefly describes the dimension-8 Effective Field Theory (EFT) approach for the aQGC $ZZ\gamma\gamma$ and $Z\gamma\gamma\gamma$. The study of the $Z\gamma\gamma$ signal and the sensitivity on the $f_{T,j}/\Lambda^4$ couplings at $95\%$ confidence level (C.L.) at the CLIC are discussed in Section III. Finally, we summarize our conclusions in Section IV.

\section{Dimension-8 effective operators for the $Z\gamma\gamma\gamma$ and $ZZ\gamma\gamma$ aQGC}

With the discovery of the Higgs boson and the measurements of its properties at the LHC, the general features of the
SM electroweak theory have been confirmed experimentally \cite{Dawson-arXiv:1808.01324}. Measurements of the Higgs
couplings and production rates agree with SM predictions, and there is no indication of the existence
of any new particle at the TeV scale \cite{Data2020}. Going forward, the task is to make comparisons between theory
and data at the few percent levels. This requires not only high-luminosity LHC running but also improved theoretical
calculations.

$Z\gamma\gamma$ triple production is an example of a process whose properties are restricted by LEP-II measurements
\cite{PLB540-2002,PRD70-2004}, yet still, provide relevant information about the $Z\gamma\gamma\gamma$ and $ZZ\gamma\gamma$
couplings from LHC data. The production of $Z\gamma\gamma$ triple provides a sensitive test of the electroweak gauge structure
of the SM and checks for non-SM aQGC.

The aQGCs induced processes can be classed into the vector boson scattering (VBS) processes and the tri-boson productions. The VBS processes are generally more sensitive to the aQGCs than the tri-boson processes \cite{Sirunyan-JHEP10-2017,ATLAS1,ATLAS2,CMS1}. At the $e^{-}e^{+}$ collider, the gauge bosons in the final states of VBS processes are associated with very energetic electrons, positrons, or neutrinos in the forward region with respect to the beam. The tri-boson processes are normally suppressed by $1/s$ in the s-channel propagator. Nevertheless, it is anticipated that the tri-boson processes without forward objects carrying away energies can also be sensitive to the aQGCs induced by dimension-8 operators with four field strength tensors.

BSM physics effects in $Z\gamma\gamma$ production can be studied using effective Lagrangian techniques where the new physics
is parameterized as an operator expansion in inverse powers of a high scale, $\Lambda$. In this regard, the extension of the
effective SM-Lagrangian is done by introducing additional dimension-8 operators for the aQGC. The formalism of the aQGC has been
widely discussed in the literature \cite{Eboli1,Eboli2,Eboli4,Eboli,Degrande,Eboli3,Eboli-PRD101-2020}. The effective Lagrangian
that gives all quartic couplings is given as follows

\begin{equation}
{\cal L}_{eff}= {\cal L}_{SM} +\sum_{k=0}^1\frac{f_{S, k}}{\Lambda^4}O_{S, k} +\sum_{i=0}^{7}\frac{f_{M, i}}{\Lambda^4}O_{M, i}+\sum_{j=0,1,2,5,6,7,8,9}^{}\frac{f_{T, j}}{\Lambda^4}O_{T, j},
\end{equation}

\noindent where each $O_{S, k}$, $O_{M, i}$ and $O_{T, j}$ is a gauge-invariant operator of dimension-8 and $\frac{f_{S, k}}{\Lambda^4}$,
$\frac{f_{M, i}}{\Lambda^4}$ and $\frac{f_{T, j}}{\Lambda^4}$ are the corresponding effective coefficient. According to the structure
of the operator, there are three classes of genuine aQGC operators as shown in Eq. (3) \cite{Degrande, Eboli3}.

The dimension-8 operators given by Eq. (3) contain contributions of the Higgs boson field ($\frac{f_{S, k}}{\Lambda^4} O_{S, k}$), Higgs-Gauge
boson field ($\frac{f_{M, i}}{\Lambda^4}O_{M, i}$), and Gauge boson field ($\frac{f_{T, j}}{\Lambda^4} O_{T, j}$), corresponding to the
occurrence of the building blocks above. The corresponding genuine aQGC operators of these contributions are presented below.   \\

$\bullet$ Contribution of operators with covariant derivatives only (Scalar field):

\begin{eqnarray}
O_{S, 0}&=&[(D_\mu\Phi)^\dagger (D_\nu\Phi)]\times [(D^\mu\Phi)^\dagger (D^\nu\Phi)],  \\
O_{S, 1}&=&[(D_\mu\Phi)^\dagger (D^\mu\Phi)]\times [(D_\nu\Phi)^\dagger (D^\nu\Phi)],  \\
O_{S, 2}&=&[(D_\mu\Phi)^\dagger (D_\nu\Phi)]\times [(D^\nu\Phi)^\dagger (D^\mu\Phi)].
\end{eqnarray}

$\bullet$ Contribution of operators with Gauge boson field strength tensors only (Tensor field):

\begin{eqnarray}
O_{T, 0}&=&Tr[W_{\mu\nu} W^{\mu\nu}]\times Tr[W_{\alpha\beta}W^{\alpha\beta}],  \\
O_{T, 1}&=&Tr[W_{\alpha\nu} W^{\mu\beta}]\times Tr[W_{\mu\beta}W^{\alpha\nu}],  \\
O_{T, 2}&=&Tr[W_{\alpha\mu} W^{\mu\beta}]\times Tr[W_{\beta\nu}W^{\nu\alpha}],  \\
O_{T, 5}&=&Tr[W_{\mu\nu} W^{\mu\nu}]\times B_{\alpha\beta}B^{\alpha\beta},  \\
O_{T, 6}&=&Tr[W_{\alpha\nu} W^{\mu\beta}]\times B_{\mu\beta}B^{\alpha\nu},  \\
O_{T, 7}&=&Tr[W_{\alpha\mu} W^{\mu\beta}]\times B_{\beta\nu}B^{\nu\alpha},  \\
O_{T, 8}&=&B_{\mu\nu} B^{\mu\nu}B_{\alpha\beta}B^{\alpha\beta},  \\
O_{T, 9}&=&B_{\alpha\mu} B^{\mu\beta}B_{\beta\nu}B^{\nu\alpha}.
\end{eqnarray}

$\bullet$ Contribution of both operators with covariant derivatives and field strength tensors (Mixed field):

\begin{eqnarray}
O_{M, 0}&=&Tr[W_{\mu\nu} W^{\mu\nu}]\times [(D_\beta\Phi)^\dagger (D^\beta\Phi)],  \\
O_{M, 1}&=&Tr[W_{\mu\nu} W^{\nu\beta}]\times [(D_\beta\Phi)^\dagger (D^\mu\Phi)],  \\
O_{M, 2}&=&[B_{\mu\nu} B^{\mu\nu}]\times [(D_\beta\Phi)^\dagger (D^\beta\Phi)],  \\
O_{M, 3}&=&[B_{\mu\nu} B^{\nu\beta}]\times [(D_\beta\Phi)^\dagger (D^\mu\Phi)],  \\
O_{M, 4}&=&[(D_\mu\Phi)^\dagger W_{\beta\nu} (D^\mu\Phi)]\times B^{\beta\nu},  \\
O_{M, 5}&=&[(D_\mu\Phi)^\dagger W_{\beta\nu} (D^\nu\Phi)]\times B^{\beta\mu} + h.c. ,  \\
O_{M, 7}&=&[(D_\mu\Phi)^\dagger W_{\beta\nu} W^{\beta\mu} (D^\nu\Phi)].
\end{eqnarray}

\noindent In Eqs. (4)-(22), the different operators are classified regarding Higgs vs. gauge boson field content.
The subscripts $S$, $T$, $M$ correspond to scalar (or longitudinal), $T$ transversal, and $M$ mixed. These correspond
to covariant derivatives of the Higgs field for the longitudinal part and field strength tensors for the transversal
part, respectively. It is worth mentioning that the photon appears only in the transverse part because there is no direct
interaction with the scalar field. In the set of genuine aQGC operators given in Eqs. (4)-(22), $\Phi$ stands for the Higgs
doublet, and the covariant derivatives of the Higgs field is given by $D_\mu\Phi=(\partial_\mu + igW^j_\mu \frac{\sigma^j}{2}
+ \frac{i}{2}g'B_\mu )\Phi$, and $\sigma^j (j=1,2,3)$ represent the Pauli matrices, while $W^{\mu\nu}$ and $B^{\mu\nu}$ are
the gauge field strength tensors for $SU(2)_L$ and $U(1)_Y$.

As mentioned above, in this work, we are studying the operators that lead to aQGC without aTGC associated with them, that is, without
an aTGC counterpart. For this reason, we focus on the dimension-8 aQGC operators, as shown in Table I. This table presents a list
of all the genuine aQGC modified by dimension-8 operators. In particular, the $Z\gamma\gamma$ channel is especially sensitive
to the $f_{T,0}/\Lambda^4$, $f_{T,1}/\Lambda^4$, $f_{T,2}/\Lambda^4$, $f_{T,5}/\Lambda^4$, $f_{T,6}/\Lambda^4$, $f_{T,7}/\Lambda^4$,
$f_{T,8}/\Lambda^4$, and $f_{T,9}/\Lambda^4$ operators. For this reason, we consider only these operators in our study. In addition,
the limits obtained for the $f_{T,8}/\Lambda^4$ and $f_{T,9}/\Lambda^4$ coupling parameters are of interest because they can be extracted
only by studying the production of electroweak neutral bosons. As can be seen from Eqs. (7-9) in the Ref \cite{PRD104-2021}, the contributions of the $O_{T,0,1}$ operators to the examined process including the anomalous $ZZ\gamma\gamma$ and $Z\gamma\gamma\gamma$ couplings are the same. Therefore, we will only calculate the $f_{T,0}/\Lambda^4$  coupling in this study.


\begin{table}{H}
\caption{The aQGC altered with dimension-8 operators are shown with X.}
\begin{center}
\begin{tabular}{|l|c|c|c|c|c|c|c|c|c|}
\hline
& $WWWW$ & $WWZZ$ & $ZZZZ$ & $WW\gamma Z$ & $WW\gamma \gamma$ & $ZZZ\gamma$ & $ZZ\gamma \gamma$ & $Z \gamma\gamma\gamma$ & $\gamma\gamma\gamma\gamma$ \\
\hline
\cline{1-10}
$O_{S0}$, $O_{S1}$                     & X & X & X &   &   &   &   &   &    \\
$O_{M0}$, $O_{M1}$, $O_{M7}$ & X & X & X & X & X & X & X &   &    \\
$O_{M2}$, $O_{M3}$, $O_{M4}$, $O_{M5}$ &   & X & X & X & X & X & X &   &    \\
$O_{T0}$, $O_{T1}$, $O_{T2}$           & X & X & X & X & X & X & X & X & X  \\
$O_{T5}$, $O_{T6}$, $O_{T7}$           &   & X & X & X & X & X & X & X & X  \\
$O_{T8}$, $O_{T9}$                     &   &   & X &   &   & X & X & X & X  \\
\hline
\end{tabular}
\end{center}
\end{table}

An advantage of the EFT formulation is that one can obtain reliable bounds on the corresponding effective coefficient
$\frac{f_{M, i}}{\Lambda^4}$ and $\frac{f_{T, j}}{\Lambda^4}$. These bounds are obtained from general considerations
and are verified in all models where calculations have been performed.

\section{Electroweak $Z\gamma\gamma$ production at the CLIC and sensitivity on aQGC}

\subsection{Sensitivity of the cross-section of the $e^+ e^-  \to Z\gamma\gamma$ signal}

The total cross-section of the process $e^+ e^- \to Z\gamma \gamma \to l^{+}l^{-}\gamma\gamma$ where $Z$ boson subsequently
decays to $e^+e^-$ or $\mu^+\mu^-$ pairs, receives contributions from five Feynman diagrams at the tree-level, as shown in
Fig. 1, plus three additional ones with crossed photon lines, respectively. The first two diagrams give the anomalous
contribution for $Z\gamma\gamma\gamma$ and $ZZ\gamma\gamma$ vertices, while the remaining diagrams give the contribution
of the SM backgrounds.

The total cross-section of the process $e^+e^- \to Z\gamma\gamma$ can be expressed as the sum of the SM component
and the EFT contributions:

\begin{eqnarray}
\sigma_{Tot}\Biggl( \sqrt{s}, \frac{f_{M,i}}{\Lambda^{4}}, \frac{f_{T,j}}{\Lambda^{4}}\Biggr)
&=& \sigma_{SM}( \sqrt{s} )
+\sigma_{(INT)}\Biggl( \sqrt{s}, \frac{f_{M,i}}{\Lambda^{4}}, \frac{f_{T,j}}{\Lambda^{4}}\Biggr)   \nonumber\\
&+& \sigma_{NP}\Biggl(\sqrt{s}, \frac{f^2_{M,i}}{\Lambda^{8}}, \frac{f^2_{T,j}}{\Lambda^{8}} \Biggr)
+ \sigma_{(MIX)}\Biggl(\sqrt{s}, \frac{f_{M,i}}{\Lambda^{4}} \frac{f_{T,j}}{\Lambda^{4}} \Biggr), \nonumber\\
i= 0-7  \hspace{3mm} \mbox{and}  \hspace{3mm} j= 0-2, 5-9,    \nonumber\\
\end{eqnarray}

\noindent where $\sigma_{SM}$ is the SM prediction, $\sigma_{(INT)}$ is the interference term between SM and the EFT operators,
$\sigma_{NP}$ is the pure EFT contribution (quadratic term) and $\sigma_{(MIX)}$ is the interference between the EFT coefficients
(cross-terms). It is worth mentioning that we do not consider a global analysis for our analysis. In this study, we assume that
the interference term between the EFT operators is zero. Therefore, we take a Wilson coefficient different from zero at a time.

In this study, the signal involves nonzero the aQGC, the SM contribution, and its interference. Here, SM represents
the SM background process of the same final state as the signal in our analysis. Also, we consider that there are also other
backgrounds such as $e^+e^- \to \tau^+\tau^-\gamma\gamma$ and $e^+e^- \to l^+l^-\gamma\gamma\gamma$. Therefore, in this work,
we study the features of the total cross-section $\sigma_{Tot}\biggl(\sqrt{s},\frac{f_{T,j}}{\Lambda^{4}}\biggr)$, the number
of events of the signals, SM and relevant backgrounds processes, as well as the sensitivity on the anomalous $f_{Tj}/\Lambda^4$
couplings using the MadGraph5\_aMC@NLO \cite{MadGraph} toolkit, where the operators given in Eqs. (7)-(14) are implemented into
MadGraph5\_aMC@NLO through Feynrules package \cite{AAlloul} as a Universal FeynRules Output (UFO) module \cite{CDegrande}.

To carry out our study, we apply the kinematic selection cuts given in Table II to suppress the backgrounds and to optimize the signal
sensitivity of the process $e^+e^- \to Z\gamma\gamma \to l^{+} l^{-}\gamma\gamma $ (Here, $l = e, \mu$). $p_{T}^{l}$ and $p_{T}^{\gamma}$
are the transverse momentums of the final state leptons, and the photons, $\eta^{l}$, and $\eta^{\gamma}$ are the pseudorapidities
of the leptons and the photons. In addition, $p^{\gamma}_{T}$ represents the scalar sum of the transverse momentum of all photons.
Here, we consider $p_{T}^{l} > 25$ GeV and  $\eta^{l}< 2.5$ with tagged Cut-1, and $\eta^{\gamma}< 2.5$
with tagged Cut-2.  To have well-separated photons and charged leptons in the phase space, we require angular separations $(\Delta R =( (\Delta \phi)^2+ (\Delta \eta)^2)^{1/2}$ between charged lepton and photon in the pseudorapidity-azimuthal angle plane that are $\Delta R(\gamma, \gamma) > 0.4$, $\Delta R(l^+, l^-) > 0.4$, $\Delta R(\gamma, l) > 0.4$ with tagged Cut-3. Also, we use the invariant mass cut for charged leptons in the final state is applied $80 < M_{l^{+}l^{-}} < 100$ GeV with tagged Cut-4.

Production of $e^+ e^- \to Z\gamma\gamma$ events with a pair of photons in the final state, with large transverse momentum $p_{T}^{\gamma}$ is especially sensitive for new physics BSM. Photons in the final state of the process $e^+ e^- \to Z\gamma\gamma$ at the CLIC have the advantage of  being identifiable with high purity and efficiency. In this sense, it is known that the high dimensional operators could affect $p_{T}^{\gamma}$ photon transverse momentum, especially at the region with large $p_{T}^{\gamma}$ values, which is very useful for distinguishing between signal and background events.

Since requiring the high transverse momentum photon eliminates the fake backgrounds, the number of expected events as a function of $p_{T}^{\gamma}$ photon transverse momentum for the $e^+ e^- \to Z\gamma \gamma$ signal and backgrounds at the CLIC with $\sqrt{s} = 3$ TeV, $P_{e^-}=0\%$, are presented in Fig. 2, to identify the region that is sensitive to the aQGC. As can be seen from this figure, above $p_{T}^{\gamma}>300$ GeV, the signal begins to separation from the backgrounds ($p_{T}^{\gamma}>300$ GeV with tagged Cut-5). It is appropriate to mention that since the other polarization options of the CLIC have the same behaviors, they will not be given separately here.

In Table III, we give the quadratic functions that exhibit the cross-section dependence as a function of the $\lambda=f_{T,j}/\Lambda^{4}$ coefficients for the $Z \gamma \gamma$ signal at the CLIC. Here the given quadratic function consists of the SM term, the interference term, and the quadratic term arising from the pure EFT contribution. These functions are composed after applying the selected cuts for the $Z\gamma \gamma$ signal at the CLIC.


\begin{table}{H}
\caption{Particle-level selections cuts for the $Z\gamma\gamma$ signal at the ILC and the CLIC.}
\begin{tabular}{|c|c|c}
\hline
Kinematic cuts &       \\
\hline
\hline
Cut-1   & \, $p^l_T>25$ GeV , $|\eta^{l}| < 2.5$    \\
\hline
Cut-2   &  $|\eta^{\gamma}| < 2.5$   \\
\hline
Cut-3  & \multicolumn{1}{|c|}{$\Delta R(\gamma,\gamma) > 0.4$ , $\Delta R(l^+, l^-) > 0.4$ , $\Delta R(\gamma,l) > 0.4$ }\\
\hline
Cut-4   & 80 GeV $< M_{l^+l^-} < 100$ GeV \\
\hline
Cut-5   & $p^\gamma_T > 300$ GeV\\
\hline
\end{tabular}
\end{table}


\begin{table}
\caption{The total cross-section as a function of the $f_{T,j}/\Lambda^{4}= \lambda$ coefficients for the  $Z \gamma \gamma$ signal at the CLIC and the ILC. }
\begin{tabular}{|c|c|c}
\hline
Parameter $\lambda$ ($\rm GeV^{-4}$) & Quadratic functions for CLIC(pb)\\
\hline
\hline
$f_{T0}/\Lambda^{4}$   &$ \sigma(\lambda)=3.77 \times 10^{-5}+9.70\times 10^{5}\lambda + 2.08\times
10^{20} \lambda^{2}$    \\
\hline
$f_{T2}/\Lambda^{4}$  &  $ \sigma(\lambda)=3.77 \times 10^{-5}-5.53\times 10^{5}\lambda + 5.11\times 10^{19} \lambda^{2}$     \\
\hline
$f_{T5}/\Lambda^{4}$   &  $ \sigma(\lambda)=3.77 \times 10^{-5}+5.22\times 10^{6}\lambda + 9.67\times 10^{20} \lambda^{2}$     \\
\hline
$f_{T6}/\Lambda^{4}$   &  $ \sigma(\lambda)=3.77 \times 10^{-5}-9.91\times 10^{5}\lambda + 5.93\times 10^{20} \lambda^{2}$     \\
\hline
$f_{T7}/\Lambda^{4}$  &   $ \sigma(\lambda)=3.77 \times 10^{-5}+3.79\times 10^{6}\lambda + 1.27\times 10^{20} \lambda^{2}$      \\
\hline
$f_{T8}/\Lambda^{4}$   &  $ \sigma(\lambda)=3.77 \times 10^{-5}-1.12\times 10^{7}\lambda + 1.64\times 10^{22} \lambda^{2}$       \\
\hline
$f_{T9}/\Lambda^{4}$ &   $ \sigma(\lambda)=3.77 \times 10^{-5}+1.43\times 10^{7}\lambda + 4.07\times 10^{21} \lambda^{2}$        \\
\hline
\hline
Parameter $\lambda$ ($\rm GeV^{-4}$) & Quadratic functions for ILC (pb)\\
\hline
\hline
$f_{T0}/\Lambda^{4}$   &$ \sigma(\lambda)=8.63 \times 10^{-5}+1.25\times 10^{6}\lambda + 2.96\times 10^{17} \lambda^{2}$    \\
\hline
$f_{T2}/\Lambda^{4}$  &  $ \sigma(\lambda)=8.63 \times 10^{-5}+4.37\times 10^{5}\lambda + 7.91\times 10^{16} \lambda^{2}$     \\
\hline
$f_{T5}/\Lambda^{4}$   &  $ \sigma(\lambda)=8.63 \times 10^{-5}-2.09\times 10^{5}\lambda + 1.51\times 10^{18} \lambda^{2}$     \\
\hline
$f_{T6}/\Lambda^{4}$   &  $ \sigma(\lambda)=8.63 \times 10^{-5}-3.58\times 10^{4}\lambda + 8.78\times 10^{17} \lambda^{2}$     \\
\hline
$f_{T7}/\Lambda^{4}$  &   $ \sigma(\lambda)=8.63 \times 10^{-5}+6.04\times 10^{5}\lambda + 1.98\times 10^{17} \lambda^{2}$      \\
\hline
$f_{T8}/\Lambda^{4}$   &  $ \sigma(\lambda)=8.63 \times 10^{-5}+5.10\times 10^{6}\lambda + 2.35\times 10^{19} \lambda^{2}$       \\
\hline
$f_{T9}/\Lambda^{4}$ &   $ \sigma(\lambda)=8.63 \times 10^{-5}-2.41\times 10^{6}\lambda + 6.28\times 10^{18} \lambda^{2}$        \\
\hline
\end{tabular}
\end{table}

In Tables IV-VI, the number of events of the $e^+ e^- \to Z\gamma \gamma$ signal is given to visualize the effects of the kinematic cuts used in the analysis of the signal and the relevant backgrounds for three different polarization states of the electron beams. For illustrative purposes, we selected $f_{T,0}/\Lambda^4=f_{T,7}/\Lambda^4=f_{T,9}/\Lambda^4 =1$ TeV$^{-4}$ and ${\cal L}=500$ fb$^{-1}$. As can be seen in these tables that $\tau\tau \gamma\gamma$ and $\gamma\gamma\gamma l^+ l^-$ backgrounds are reduced more than the SM background and the signal. Also, the number of expected events of the $e^+ e^- \to Z\gamma \gamma$ signal is a factor of 2 - 18 greater than the corresponding ones for the SM and the relevant background. It is worth mentioning that
Cut-0 in Tables IV-VI refers to the default cuts which include the minimal values of the objects to handle the inferred divergences.


The next observable that we study is the total cross-sections for the process $e^+ e^- \to Z\gamma \gamma\to l^{+}l^{-}\gamma\gamma$ as a function of the
Wilson coefficients $f_{Tj}/\Lambda^4$ (see Table III), which are shown in Figs. 3-8.  Here, we consider unpolarized and polarized beams of electrons, that is $P_{e^-}=-80\%, 0\%, 80\%$. Also, each operator coefficient is scanned independently, with all other operator's coefficients set to zero.  The results demonstrate a clear dependence on the cross-section of the process $e^+ e^- \to Z\gamma \gamma \to l^{+}l^{-}\gamma\gamma$ concerning the aQGC. As seen from these figures, the sensitivity of the coefficients $f_{T,8}/\Lambda^4$ and $f_{T,9}/\Lambda^4$ is more important than the other aQGC. Thus, we anticipate obtaining better limits on the coefficients $f_{T,8}/\Lambda^4$ and $f_{T,9}/\Lambda^4$ via the process $e^+ e^- \to Z\gamma \gamma\to l^{+}l^{-}\gamma\gamma$ at the CLIC.


The Initial State Radiation (ISR) is an essential issue in high-energy processes, especially for lepton colliders. ISR significantly
affects both the SM and anomalous coupling cases. Also, for the linear colliders, the energy loss due to Beamstrahlung
during the collision of the incoming lepton beams are expected to substantially influence the effective center-of-mass energy
distribution of the colliding particles.

In this paper, ISR and Beamstrahlung effects are implemented in MadGraph5\_aMC@NLO \cite{Frixione-arXiv:2108.10261}. We only consider the effects of ISR and Beamstrahlung on the cross-sections of the signal and backgrounds for the process $e^+e^- \to Z\gamma\gamma$, while obtaining the best limit values on the anomalous couplings.

In Figs. 6-8, we give the total cross-sections for the process $e^+e^-\to Z\gamma\gamma$ considering ISR and Beamstrahlung effects in terms of the aQGC $f_ {T,j}/\Lambda^4$ for the CLIC with $\sqrt{s}=3$ TeV.

In Refs. \cite{PRD93-2016, JHEP10-2017, JHEP10-2021}, the ATLAS and CMS Collaborations at the LHC report evidence on the production $pp \to Z\gamma \gamma$, which is accessible for the first time with 8 TeV and 20.3 fb$^{-1}$ data set and more recently with $\sqrt{s} = 13$ TeV and ${\cal L}= 137$ fb$^{-1}$. In these papers, the measured cross-section is used to set limits on aQGC $Z\gamma\gamma\gamma$ and $ZZ\gamma\gamma$. Its bounds on the aQGC $Z\gamma\gamma\gamma$ and $ZZ\gamma\gamma$ are obtained from selecting the most relevant systematic uncertainties, which are of the order of $1-10\%$. The dominant systematic uncertainties come from estimating the backgrounds 9\%, the systematic uncertainty coming from the jet-photon misidentification $5-6\%$, integrated luminosity $1-3\%$ and photon efficiencies $3-6\%$. For more details, see Tables 1 and 4 from these references. In addition,
the L3 and OPAL Collaborations \cite{PLB540-2002,PRD70-2004,PLB478-2000} report measurements of the $e^+ e^- \to Z\gamma\gamma$
cross-section and determination of quartic gauge boson couplings at LEP using a total integrated luminosity of 231 $\rm pb^{-1}$
collected with the L3 detector at center-of-mass energies of 182.7 GeV and 188.7 GeV \cite{PLB540-2002}. The systematic uncertainties
are listed in Table 2 \cite{PLB478-2000}. Also, there are phenomenological studies in the literature for the future CLIC in which moderate values for the
systematic uncertainties of $0\%$, $3\%$, $5\%$, and $10\%$ \cite{PRD98-2018,PRD98-015017-2018} are considered.
Based on the systematic uncertainties reported in Refs. \cite{PRD93-2016,JHEP10-2017,JHEP10-2021,PLB540-2002,PRD70-2004,PLB478-2000,PRD98-2018,PRD98-015017-2018}, and in order to obtain the sensitivity on the anomalous $Z\gamma\gamma\gamma$ and $ZZ\gamma\gamma$ couplings, we choose the systematic uncertainties of $\delta_{sys} =$ $0\%$, $3\%$, and $5\%$, respectively. In the next subsection, we present our results on the expected sensitivity of the Wilson coefficients $\frac{f_{T, j}}{\Lambda^4}$.

\begin{table}{H}
\caption{Number of expected events at the CLIC for the process $e^+e^- \to Z\gamma\gamma\to l^+l^-\gamma\gamma$, SM and relevant backgrounds processes
         after applied cuts given in Table II, and $P_{e^-}=-80\%$.}
\begin{tabular}{|c|c|c|c|c|c}
\hline
\multicolumn{5}{|c|}{Polarized beams, $P_{e^-}=-80\%$ }\\
\hline
Kinematic cuts & Signal ($f_{T0}/\Lambda^{4}$ = 1 TeV$^{-4}$) & SM & $e^{-}e^{+}\to\tau\tau\gamma \gamma$ & $e^{-}e^{+}\to\gamma\gamma\gamma l^+ l^-$ \\
\hline
\hline
Cut-0 &395 & 86 & 125 & 77  \\
\hline
Cut-1  & 395 & 86 & 0 &  77    \\
\hline
Cut-2  & 395 & 86 & 0 &   77   \\
\hline
Cut-3  & 395 & 86 & 0 &  77 \\
\hline
Cut-4  & 367 &80 & 0 &  5 \\
\hline
Cut-5  &279 & 27 & 0 & 0  \\
\hline
 & Signal($f_{T7}/\Lambda^{4}$ = 1 TeV$^{-4}$) & SM & $e^{-}e^{+}\to\tau\tau\gamma \gamma$ & $e^{-}e^{+}\to\gamma\gamma\gamma l^+ l^-$ \\
\hline
\hline
Cut-0 &196 & 86 & 125 & 77  \\
\hline
Cut-1  & 196 & 86 & 0 &  77   \\
\hline
Cut-2  & 196 & 86 & 0 &   77   \\
\hline
Cut-3  & 196 & 86 & 0 &   77 \\
\hline
Cut-4  & 182 & 80 & 0 & 5 \\
\hline
Cut-5  & 119 & 27 & 0 &  0  \\
\hline
\hline
 & Signal($f_{T9}/\Lambda^{4}$ = 1 TeV$^{-4}$) & SM & $e^{-}e^{+}\to\tau\tau\gamma \gamma$ & $e^{-}e^{+}\to\gamma\gamma\gamma l^+ l^-$ \\
\hline
\hline
Cut-0 &1600 & 86 & 125 & 77  \\
\hline
Cut-1  & 1600 & 86 & 0 &  77   \\
\hline
Cut-2  & 1600 & 86 & 0 &   77   \\
\hline
Cut-3  & 1600 & 86 & 0 &   77 \\
\hline
Cut-4  & 1490 & 80 & 0 & 5 \\
\hline
Cut-5  & 968 & 27 & 0 &  0  \\
\hline
\end{tabular}
\end{table}

\begin{table}{H}
\caption{Same as in Table IV, but for $P_{e^-}=0\%$.
}
\begin{tabular}{|c|c|c|c|c|c}
\hline
\multicolumn{5}{|c|}{Polarized beams, $P_{e^-}=0\%$ }\\
\hline
Kinematic cuts & Signal ($f_{T0}/\Lambda^{4}$ = 1 TeV$^{-4}$) & SM & $e^{-}e^{+}\to\tau\tau\gamma \gamma$ & $e^{-}e^{+}\to\gamma\gamma\gamma l^+ l^-$ \\
\hline
\hline
Cut-0 &249 & 80 & 118 & 72  \\
\hline
Cut-1  & 249 & 80 & 0 & 72     \\
\hline
Cut-2  & 249 & 80 & 0 & 72     \\
\hline
Cut-3  & 249 & 80 & 0 & 72  \\
\hline
Cut-4  & 231 &75 & 0 & 5  \\
\hline
Cut-5  &165 & 25 & 0 & 0  \\
\hline
 & Signal($f_{T7}/\Lambda^{4}$ = 1 TeV$^{-4}$) & SM & $e^{-}e^{+}\to\tau\tau\gamma \gamma$ & $e^{-}e^{+}\to\gamma\gamma\gamma l^+ l^-$ \\
\hline
\hline
Cut-0 &170 & 80 & 118 & 72  \\
\hline
Cut-1  & 170 & 80 & 0 & 72    \\
\hline
Cut-2  & 170 & 80 & 0 & 72     \\
\hline
Cut-3  & 170 & 80 & 0 & 72   \\
\hline
Cut-4  & 158 & 75 & 0 & 5 \\
\hline
Cut-5  & 101 & 25 & 0 & 0   \\
\hline
\hline
 & Signal($f_{T9}/\Lambda^{4}$ = 1 TeV$^{-4}$) & SM & $e^{-}e^{+}\to\tau\tau\gamma \gamma$ & $e^{-}e^{+}\to\gamma\gamma\gamma l^+ l^-$ \\
\hline
\hline
Cut-0 &3130 & 80 & 118 & 72  \\
\hline
Cut-1  & 3130 & 80 & 0 & 72    \\
\hline
Cut-2  & 3130 & 80 & 0 & 72     \\
\hline
Cut-3  & 3130 & 80 & 0 & 72   \\
\hline
Cut-4  & 2902 & 75 & 0 & 5 \\
\hline
Cut-5  & 2540 & 25 & 0 & 0   \\
\hline
\end{tabular}
\end{table}

\begin{table}{H}
\caption{Same as in Table IV, but for $P_{e^-}=80\%$.
}
\begin{tabular}{|c|c|c|c|c|c}
\hline
\multicolumn{5}{|c|}{ Polarized beams, $P_{e^-}=80\%$ }\\
\hline
Kinematic cuts & Signal ($f_{T0}/\Lambda^{4}$ = 1 TeV$^{-4}$) & SM & $e^{-}e^{+}\to\tau\tau\gamma \gamma$ & $e^{-}e^{+}\to\gamma\gamma\gamma l^+ l^-$ \\
\hline
\hline
Cut-0 &105 & 70 & 110 & 62  \\
\hline
Cut-1  & 105 & 70 & 0 &  62    \\
\hline
Cut-2  & 105 & 70 & 0 &   62   \\
\hline
Cut-3  & 105 & 70 & 0 &  62 \\
\hline
Cut-4  & 98 &65 & 0 &  3 \\
\hline
Cut-5  & 50 & 21 & 0 & 0  \\
\hline
 & Signal($f_{T7}/\Lambda^{4}$ = 1 TeV$^{-4}$) & SM & $e^{-}e^{+}\to\tau\tau\gamma \gamma$ & $e^{-}e^{+}\to\gamma\gamma\gamma l^+ l^-$ \\
\hline
\hline
Cut-0 &145 & 70 & 110 & 62  \\
\hline
Cut-1  & 145 & 70 & 0 &  62   \\
\hline
Cut-2  & 145 & 70 & 0 &   62   \\
\hline
Cut-3  & 145 & 70 & 0 &   62 \\
\hline
Cut-4  & 135 & 65 & 0 & 3 \\
\hline
Cut-5  & 85 & 21 & 0 &  0  \\
\hline
\hline
 & Signal($f_{T9}/\Lambda^{4}$ = 1 TeV$^{-4}$) & SM & $e^{-}e^{+}\to\tau\tau\gamma \gamma$ & $e^{-}e^{+}\to\gamma\gamma\gamma l^+ l^-$ \\
\hline
\hline
Cut-0 &4700 & 70 & 110 & 62  \\
\hline
Cut-1  & 4700 & 70 & 0 &  62   \\
\hline
Cut-2  & 4700 & 70 & 0 &   62   \\
\hline
Cut-3  & 4700 & 70 & 0 &   62 \\
\hline
Cut-4  & 4358 & 65 & 0 & 3 \\
\hline
Cut-5  & 3855 & 21 & 0 &  0  \\
\hline
\end{tabular}
\end{table}

\subsection{Expected sensitivity on the Wilson coefficients $\frac{f_{T, j}}{\Lambda^4}$ at the CLIC}

The presence of new physics characterized by the parameters $f_ {T,j}/\Lambda^4$, $j=0-2, 5-9$, may be quantified by a simple method that varies the parameters and is based on the chi-squared distribution:

\begin{equation}
\chi^2(f_{T,j}/\Lambda^4)=\Biggl(\frac{\sigma_{SM}(\sqrt{s})-\sigma_{BSM}(\sqrt{s}, f_{T,j}/\Lambda^4)}
{\sigma_{SM}(\sqrt{s})\sqrt{(\delta_{st})^2 + (\delta_{sys})^2}}\Biggr)^2,
\end{equation}

\noindent where $\sigma_{SM}(\sqrt{s})$ is the cross-section in the SM and $\sigma_{BSM}(\sqrt{s}, f_{T,j}/\Lambda^4)$
is the cross-section in the presence of BSM interactions, $\delta_{st}=\frac{1}{\sqrt{N_{SM}}}$
is the statistical error and $\delta_{sys}$ is the systematic error. The number of events is given by $N_{SM}=
{\cal L}\times \sigma_{SM}$, where ${\cal L}$ is the integrated luminosity of the CLIC.





\begin{table}{H}
\caption{Expected sensitivity at $95\%$ C.L. on the aQGC $ZZ\gamma\gamma$ and $Z\gamma\gamma\gamma$ through the process
$e^+e^- \to Z\gamma\gamma$ in the case of two decay channels at the CLIC with $P_{e^-}=0\%, -80\%, 80\%$, $\delta_{sys}=0\%, 3\%,
5\%$ are represented. Here, while any coupling is calculated, the other couplings are set to zero.}
\begin{tabular}{|c|c|c|c|c|}
\hline
$P_{e^-}$              &      & $0\%$       & $-80\%$    & $80\%$ \\
\hline
Couplings (TeV$^{-4}$) & & ${\cal L}=5$ ab$^{-1}$ & ${\cal L}=4$ ab$^{-1}$ & ${\cal L}=1$ ab$^{-1}$ \\
\hline
                      &$\delta=0\%$       &$[-1.64;1.59]\times10^{-1}$  &$[-1.33;1.28]\times10^{-1}$ &$[-5.36;5.13]\times10^{-1}$ \\
$f_{T0}/\Lambda^{4}$  &$\delta=3\%$       &$[-1.70;1.66]\times10^{-1}$  &$[-1.38;1.33]\times10^{-1}$ &$[-5.41;5.18]\times10^{-1}$ \\
                      &$\ \, \delta=5\%$ &$[-1.80;1.75]\times10^{-1}$  &$[-1.45;1.40]\times10^{-1}$ &$[-5.49;5.26]\times10^{-1}$ \\
\hline
                      &$\delta=0\%$       &$[-3.20;3.31]\times10^{-1}$  &$[-2.59;2.67]\times10^{-1}$ &$[-1.03;1.09]$ \\
$f_{T2}/\Lambda^{4}$  &$\delta=3\%$       &$[-3.33;3.43]\times10^{-1}$  &$[-2.68;2.76]\times10^{-1}$ &$[-1.04;1.09]$ \\
                      &$\ \, \delta=5\%$ &$[-3.53;3.64]\times10^{-1}$  &$[-2.83;2.91]\times10^{-1}$ &$[-1.06;1.11]$ \\
\hline
                      &$\delta=0\%$       &$[-7.77;7.23]\times10^{-2}$  &$[-7.62;7.00]\times10^{-2}$ &$[-1.25;1.23]\times10^{-1}$ \\
$f_{T5}/\Lambda^{4}$  &$\delta=3\%$       &$[-8.07;7.53]\times10^{-2}$  &$[-7.88;7.26]\times10^{-2}$ &$[-1.26;1.24]\times10^{-1}$ \\
                      &$\ \, \delta=5\%$ &$[-8.53;7.99]\times10^{-2}$  &$[-8.29;7.66]\times10^{-2}$ &$[-1.27;1.25]\times10^{-1}$ \\
\hline
                     &$\delta=0\%$       &$[-9.48;9.65]\times10^{-2}$  &$[-1.15;1.24]\times10^{-1}$ &$[-1.11;1.40]\times10^{-1}$ \\
$f_{T6}/\Lambda^{4}$ &$\delta=3\%$       &$[-0.99;1.00]\times10^{-1}$  &$[-1.19;1.28]\times10^{-1}$ &$[-1.12;1.41]\times10^{-1}$ \\
                     &$\ \, \delta=5\%$ &$[-1.05;1.06]\times10^{-1}$  &$[-1.26;1.35]\times10^{-1}$ &$[-1.14;1.43]\times10^{-1}$ \\
\hline
                      &$\delta=0\%$       &$[-2.22;1.92]\times10^{-1}$  &$[-2.39;2.15]\times10^{-1}$ &$[-3.06;2.88]\times10^{-1}$ \\
$f_{T7}/\Lambda^{4}$  &$\delta=3\%$       &$[-2.31;2.01]\times10^{-1}$  &$[-2.47;2.23]\times10^{-1}$ &$[-3.00;2.98]\times10^{-1}$ \\
                      &$\ \, \delta=5\%$ &$[-2.43;2.13]\times10^{-1}$  &$[-2.59;2.35]\times10^{-1}$ &$[-3.04;3.02]\times10^{-1}$ \\
\hline
                      &$\delta=0\%$       &$[-1.78;1.85]\times10^{-2}$  &$[-2.44;3.04]\times10^{-2}$ &  $[-2.21;2.13]\times10^{-2}$\\
$f_{T8}/\Lambda^{4}$  &$\delta=3\%$       &$[-1.85;1.92]\times10^{-2}$  &$[-2.53;3.14]\times10^{-2}$ &  $[-2.23;2.15]\times10^{-2}$\\
                      &$\ \, \delta=5\%$ &$[-1.96;2.03]\times10^{-2}$  &$[-2.68;3.29]\times10^{-2}$ &  $[-2.26;2.18]\times10^{-2}$\\
\hline
                      &$\delta=0\%$       &$[-3.83;3.48]\times10^{-2}$  &$[-5.83;5.17]\times10^{-2}$ &$[-4.39;4.35]\times10^{-2}$ \\
$f_{T9}/\Lambda^{4}$  &$\delta=3\%$       &$[-3.98;3.62]\times10^{-2}$  &$[-6.02;5.36]\times10^{-2}$ &$[-4.42;4.38]\times10^{-2}$ \\
                      &$\ \, \delta=5\%$ &$[-4.20;3.85]\times10^{-2}$  &$[-6.32;5.66]\times10^{-2}$ &$[-4.48;4.44]\times10^{-2}$ \\
\hline
\end{tabular}
\end{table}

\begin{table}{H}
\caption{Expected sensitivity at $95\%$ C.L. on the aQGC $ZZ\gamma\gamma$ and $Z\gamma\gamma\gamma$ through the process
$e^+e^- \to Z\gamma\gamma$ in the case of two decay channels considering ISR and Beamstrahlung effects at the CLIC with
$P_{e^-}=0\%, -80\%, 80\%$, $\delta_{sys}=0\%, 3\%, 5\%$ are represented. Here, while any
coupling is calculated, the other couplings are set to zero.}
\begin{tabular}{|c|c|c|c|c|}
\hline
$P_{e^-}$ & & 0\% & -80\% & 80\% \\
\hline
Couplings (TeV$^{-4}$) & & ${\cal L}=5$ ab$^{-1}$ &${\cal L}=4$ ab$^{-1}$ & ${\cal L}=1$ ab$^{-1}$ \\
\hline
                      & $\delta=0\%$      & $[-1.90;1.82]\times10^{-1}$ & $[-1.52;1.48]\times10^{-1}$ & $[-6.31;5.78]\times10^{-1}$ \\
$f_{T0}/\Lambda^{4}$  & $\delta=3\%$      & $[-1.98;1.90]\times10^{-1}$ & $[-1.58;1.54]\times10^{-1}$ & $[-6.36;5.83]\times10^{-1}$ \\
                      & $\ \, \delta=5\%$ & $[-2.10;2.01]\times10^{-1}$ & $[-1.66;1.62]\times10^{-1}$ & $[-6.44;5.91]\times10^{-1}$ \\
\hline
                      & $\delta=0\%$      & $[-3.54;3.95]\times10^{-1}$ & $[-2.93;3.12]\times10^{-1}$ & $[-1.16;1.29]$ \\
$f_{T2}/\Lambda^{4}$  & $\delta=3\%$      & $[-3.69;4.11]\times10^{-1}$ & $[-3.04;3.23]\times10^{-1}$ & $[-1.17;1.30]$ \\
                      & $\ \, \delta=5\%$ & $[-3.93;4.35]\times10^{-1}$ & $[-3.21;3.40]\times10^{-1}$ & $[-1.19;1.31]$ \\
\hline
                      & $\delta=0\%$      & $[-8.63;8.51]\times10^{-2}$ & $[-8.90;7.93]\times10^{-2}$ & $[-1.46;1.38]\times10^{-1}$ \\
$f_{T5}/\Lambda^{4}$  & $\delta=3\%$      & $[-8.98;8.87]\times10^{-2}$ & $[-9.21;8.24]\times10^{-2}$ & $[-1.47;1.39]\times10^{-1}$ \\
                      & $\ \, \delta=5\%$ & $[-9.53;9.42]\times10^{-2}$ & $[-9.69;8.72]\times10^{-2}$ & $[-1.49;1.41]\times10^{-1}$  \\
\hline
                      & $\delta=0\%$      & $[-1.05;1.15]\times10^{-1}$ & $[-1.33;1.40]\times10^{-1}$ & $[-1.35;1.52]\times10^{-1}$ \\
$f_{T6}/\Lambda^{4}$  & $\delta=3\%$      & $[-1.10;1.19]\times10^{-1}$ & $[-1.38;1.45]\times10^{-1}$ & $[-1.36;1.53]\times10^{-1}$ \\
                      & $\ \, \delta=5\%$ & $[-1.17;1.26]\times10^{-1}$ & $[-1.46;1.53]\times10^{-1}$ & $[-1.38;1.55]\times10^{-1}$ \\
\hline
                      & $\delta=0\%$      & $[-2.60;2.17]\times10^{-1}$ & $[-2.84;2.39]\times10^{-1}$ & $[-3.95;2.96]\times10^{-1}$ \\
$f_{T7}/\Lambda^{4}$  & $\delta=3\%$      & $[-2.70;2.27]\times10^{-1}$ & $[-2.94;2.48]\times10^{-1}$ & $[-3.98;2.99]\times10^{-1}$ \\
                      & $\ \, \delta=5\%$ & $[-2.85;2.42]\times10^{-1}$ & $[-3.09;2.63]\times10^{-1}$ & $[-4.02;3.03]\times10^{-1}$ \\
\hline
                      &$\delta=0\%$      & $[-1.91;2.28]\times10^{-2}$ & $[-2.71;3.60]\times10^{-2}$ & $[-2.72;2.29]\times10^{-2}$ \\
$f_{T8}/\Lambda^{4}$  &$\delta=3\%$      & $[-2.00;2.37]\times10^{-2}$ & $[-2.83;3.71]\times10^{-2}$ & $[-2.74;2.31]\times10^{-2}$ \\
                      &$\ \, \delta=5\%$ & $[-2.13;2.50]\times10^{-2}$ & $[-3.00;3.89]\times10^{-2}$ & $[-2.78;2.35]\times10^{-2}$ \\
\hline
                      &$\delta=0\%$      & $[-4.21;4.17]\times10^{-2}$ & $[-6.53;6.10]\times10^{-2}$ &$[-5.72;4.38]\times10^{-2}$ \\
$f_{T9}/\Lambda^{4}$  &$\delta=3\%$      & $[-4.38;4.34]\times10^{-2}$ & $[-6.76;6.34]\times10^{-2}$ &$[-5.76;4.42]\times10^{-2}$ \\
                      &$\ \, \delta=5\%$ & $[-4.65;4.61]\times10^{-2}$ & $[-7.13;6.70]\times10^{-2}$ &$[-5.83;4.49]\times10^{-2}$ \\
\hline
\end{tabular}
\end{table}

\begin{table}{H}
\caption{Expected sensitivity at $95\%$ C.L. on the aQGC $ZZ\gamma\gamma$ and $Z\gamma\gamma\gamma$ through the process
$e^+e^- \to Z\gamma\gamma $ in the case of two decay channels at the ILC with $\sqrt{s}=1$ TeV, ${\cal L}=8\hspace{1mm}
\rm ab^{-1}$, $\delta_{sys}=0\%, 3\%, 5\%$ and $P_{e^-}=0\%$ are represented. Here, while any
coupling is calculated, the other couplings are set to zero.}
\begin{tabular}{|c|c|c|c}
\hline
Couplings (TeV$^{-4}$) & & ${\cal L}=8\hspace{0.8mm} {\rm ab^{-1}}$ \\
\hline
\hline                                    &  $\delta=0\%$        &  $[-0.72;0.30]\times10^{1}$ \\
$f_{T0}/\Lambda^{4}$  &  $\delta=3\%$        &  $[-0.78;0.36]\times10^{1}$ \\
                                    &  $\ \, \delta=5\%$ &  $[-0.85;0.43]\times10^{1}$\\
\hline
                                    & $\delta=0\%$          & $[-1.22;0.67]\times10^{1}$\\
$f_{T2}/\Lambda^{4}$  &  $\delta=3\%$         & $[-1.33;0.78]\times10^{1}$\\
                                    &  $\ \, \delta=5\%$  &$[-1.47;0.92]\times10^{1}$\\
\hline
                                    &$\delta=0\%$          &$[-2.00;2.14]$ \\
$f_{T5}/\Lambda^{4}$  & $\delta=3\%$         & $[-2.27;2.41]$\\
                                    &$\ \, \delta=5\%$  & $[-2.59;2.73]$\\
\hline
                                    &$\delta=0\%$         & $[-2.70;2.74]$\\
$f_{T6}/\Lambda^{4}$  & $\delta=3\%$        & $[-3.04;3.08]$\\
                                    &$\ \, \delta=5\%$  & $[-3.47;3.51]$\\
\hline
                                    &$\delta=0\%$          &$[-0.74;0.44]\times10^{1}$\\
$f_{T7}/\Lambda^{4}$  & $\delta=3\%$         &$[-0.81;0.51]\times10^{1}$\\
                                    &$\ \, \delta=5\%$  &$[-0.90;0.60]\times10^{1}$\\
\hline
                                    &$\delta=0\%$           & $[-0.64;0.43]$\\
$f_{T8}/\Lambda^{4}$  &$\delta=3\%$           & $[-0.71;0.49]$\\
                                    & $\ \, \delta=5\%$   & $[-0.79;0.57]$\\
\hline
                                    &$\delta=0\%$         &$[-0.84;1.23]$\\
$f_{T9}/\Lambda^{4}$  &$\delta=3\%$          &$[-0.97;1.35]$\\
                                    &$\ \, \delta=5\%$  & $[-1.13; 1.51]$\\
\hline
\end{tabular}

\end{table}

At the CLIC, the coefficients $f_{T,j}/\Lambda^4$ can be tested at $95\%$ C.L. in the $e^+e^- \to Z\gamma\gamma$ production mode with Cut-5 given in Table II, as well as with the center-of-mass energy of $\sqrt{s}=3$ TeV, integrated luminosities of ${\cal L}=1$ ${\rm ab^{-1}}$ ($P_{e^-}=80\%$), ${\cal L}=4$ ${\rm ab^{-1}}$ ($P_{e^-}=-80\%$), ${\cal L}=5$ ${\rm ab^{-1}}$ ($P_{e^-}=0\%$) and systematic uncertainties of $\delta_{sys}=0\%, 3\%, 5\%$.

In order to quantify the expected sensitivity on the Wilson coefficients $f_{T,j}/\Lambda^4$, an advantage has been taken in this analysis of the fact that the aQGC enhances the total cross-section at high energies, as shown in Figs. 3-8.

Tables VII-VIII and Figs. 9-16, show the expected sensitivities at $95\%$ C.L. on the aQGC $ZZ\gamma\gamma$ and $Z\gamma\gamma\gamma$ through
the process $e^+e^- \to Z\gamma\gamma$ in the case of two decay channels at the CLIC with $P_{e^-}=-80\%$, $0\%$, $80\%$, $\delta_{sys} = 0\%,
3\%, 5\%$. Here, any couplings are calculated while fixing the other couplings to zero.
Our best sensitivity limits on the aQGC at the CLIC might reach up to the order of magnitude $O(10^{1}- 10^{-2})$, respectively. As can be seen from these tables and figures, the $f_ {T,8}/\Lambda^4$, and $f_ {T,9}/\Lambda^4$ couplings have the best sensitivities among the aQGC, as we expected. Our best limits on the aQGC improve much better than the current experimental limits.

As above mentioned, the current CLIC baseline staging scenario assumes 1.0 $\rm ab^{-1}$, 2.5 $\rm ab^{-1}$, and 5.0 $\rm ab^{-1}$
of luminosity at the three energy stages of $\sqrt{s}=0.38$ TeV, 1.5 TeV, and 3 TeV, respectively \cite{franc}.  The expected integrated
luminosities for the CLIC with the polarized electron beams are ${\cal L}=1$ ab$^{-1}$ ($P_{e^-}=80\%$), ${\cal L}=4$ ab$^{-1}$
($P_{e^-}=-80\%$), and ${\cal L}=5$ ab$^{-1}$ ($P_{e^-}=0\%$) \cite{franc,CLIC-1812.06018}. Starting from these scenarios, we examine
the effect of three different $P_{e^-}=-80\%$, $0\%$, $80\%$ polarizations states on the aQGC.

In Figs. 9-16, are given the comparison of the current experimental limits and projected sensitivity on $f_{T,j}/\Lambda^4$ couplings, we consider the integrated luminosities of ${\cal L} = 1$ $\rm ab^{-1}$ ($P_{e^-}=80\%$), ${\cal L} = 4$ $\rm ab^{-1}$ ($P_{e^-}=-80\%$) and ${\cal L} = 5$ $\rm ab^{-1}$ ($P_{e^-}=0\%$).


From Figs. 9-16, while our best sensitivities on $f_{T,0}/\Lambda^4$, $f_{T,1}/\Lambda^4$, $f_{T,2}/\Lambda^4$, $f_{T,5}/\Lambda^4$ and $f_{T,7}/\Lambda^4$ couplings at $P_{e^-}=-80\%$ are more restrictive than the limits obtained for the other polarizations $P_{e^-}=0\%$ and $80\%$,
$f_{T,6}/\Lambda^4$, $f_{T,8}/\Lambda^4$ and $f_{T,9}/\Lambda^4$ couplings determined at $P_{e^-}=0\%$ are better than the sensitivities derived on the aQGC for the other polarizations. This is because the $O_{T}$ operators examined in this study have different multi-boson topologies. The limits on the aQGC with systematic uncertainties of $3\%$ and $5\%$ are weaker than the limits obtained without systematic uncertainties. In Table VII, $f_{T,8}/\Lambda^4$ and
$f_{T,9}/\Lambda^4$ couplings have the best sensitivities between the aQGC, $f_{T,8}/\Lambda^4$ without systematic uncertainties are approximately 1.03 and 1.1 times better than $3\%$ and $5\%$ systematic uncertainties limits, respectively. Finally, if we compare Tables VII and VIII, the best limits obtained considering ISR and Beamstrahlung effects are between $13\%$ and $24\%$ worse than the limits derived without these effects.

A direct comparison of the projected sensitivity limits on the anomalous  $f_{T,j}/\Lambda^4$ couplings with the center-of-mass
energy and integrated luminosity  projected  for the third stage of the ILC, reported in Table IX, with the corresponding projections
for the CLIC given in Table VII, clearly show that the projections for the CLIC show a better sensitivity on anomalous $f_{T,j}/\Lambda^4$
couplings. This difference is of the order of magnitude of $O(10^{-2})$ for the CLIC compared to the ILC.

\section{RESULTS AND CONCLUSIONS}

The EFT framework represents a way to parameterize new physics phenomena, extending the SM by considering new interactions between its constituent particles. A specific class of these interactions is labeled as the aQGC. Investigating the aQGC described by the non-Abelian gauge symmetry within the framework of the SM may lead to an additional confirmation of the model and give clues to the existence of new physics. The LHC maybe provide discoveries and valuable information.

On the other hand, it is usually agreed that the clean and precise environment of the linear colliders according to the LHC is ideally suited to the examination of new physics. In addition, the use of polarized beams plays a significant role in the physics programs of future linear colliders, and since the aQGC defined through effective Lagrangian have dimension-8, these have strong energy dependence. Thus, the values of the anomalous cross-section containing the anomalous $ZZ\gamma\gamma$ and $Z\gamma\gamma\gamma$ vertices have higher than the SM cross-section. For this reason, future linear colliders such as the CLIC with high center-of-mass energy and luminosity could provide us with possible new physics signs beyond the SM. With these motivations, we perform a detailed study of the process $e^{-}e^{+}\to Z\gamma\gamma\to l^+ l^-\gamma\gamma $ at the CLIC to probe the aQGC $Z\gamma\gamma\gamma$ and $ZZ\gamma\gamma$.


In conclusion, we have shown that the study of the process $e^+e^- \to Z\gamma\gamma \to l^+ l^- \gamma\gamma$ at the CLIC can test
aQGC $f_{T,j}/\Lambda^4$ that are two orders of magnitude stronger than the existing limits from direct searches reported
by the current experimental limits. In particular, stringent sensitivities are projected
on the anomalous $f_{T,8,9}/\Lambda^4$ couplings $\frac{f_{T8}}{\Lambda^{4}}= [-1.78; 1.85] \times 10^{-2} \hspace{1mm} {\rm TeV^{-4}}$ and $\frac{f_{T9}}{\Lambda^{4}}= [-3.83; 3.48] \times 10^{-2} \hspace{1mm} {\rm TeV^{-4}}$ with $\sqrt{s}=3\hspace{0.8mm}{\rm TeV}$, ${\cal L}=5\hspace{0.8mm} \rm ab^{-1}$, $P_{e^-}=0\%$ and $\delta_{sys}=0\%$. Among the eight operators, we see that the $O_{T,8}$ and $O_{T,9}$  operators have the best sensitivities since they only contain the neutral electroweak gauge bosons. Furthermore, if a signal is observed, the comparison of the process here studied, which is sensitive only to photonic quartic operators, with the observations for processes also dependent on nonphotonic couplings, such as weak gauge boson pair production, could reveal some symmetries of the underlying dynamics. Therefore, if the experimental measurement of the aQGC were to be confirmed, then nature could have provided us with the technique and the signal appropriate to study the complete spectroscopy of the aQGC.

\vspace{1.5cm}

\begin{center}
{\bf Acknowledgements}
\end{center}

A. G. R. and M. A. H. R. thank SNI and PROFEXCE (M\'exico). The numerical calculations reported in this paper were fully performed at TUBITAK ULAKBIM, High Performance and Grid Computing Center (TRUBA resources).

\vspace{1cm}


\newpage

\begin{figure}[H]
\centerline{\scalebox{0.6}{\includegraphics{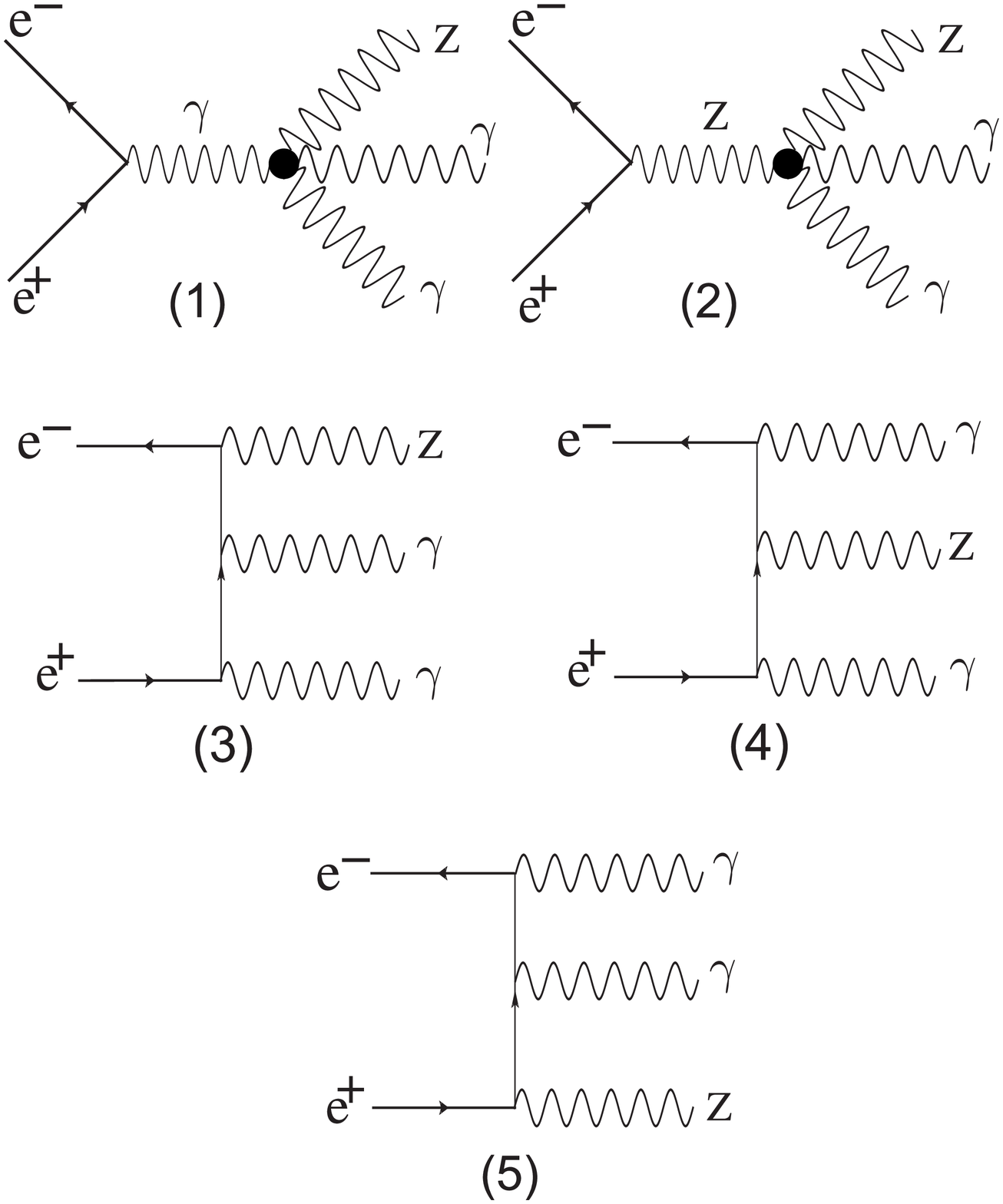}}}
\caption{ \label{fig:gamma} Feynman diagrams for the signal process $e^+e^-\to Z\gamma\gamma$
induced by the effective $ZZ\gamma\gamma$ and $Z\gamma\gamma\gamma$ vertices, plus three additional
ones with crossed photon lines. New physics (represented by a black circle) in the electroweak sector can modify the QGCs.}
\end{figure}

\begin{figure}[H]
\centerline{\scalebox{0.9}{\includegraphics{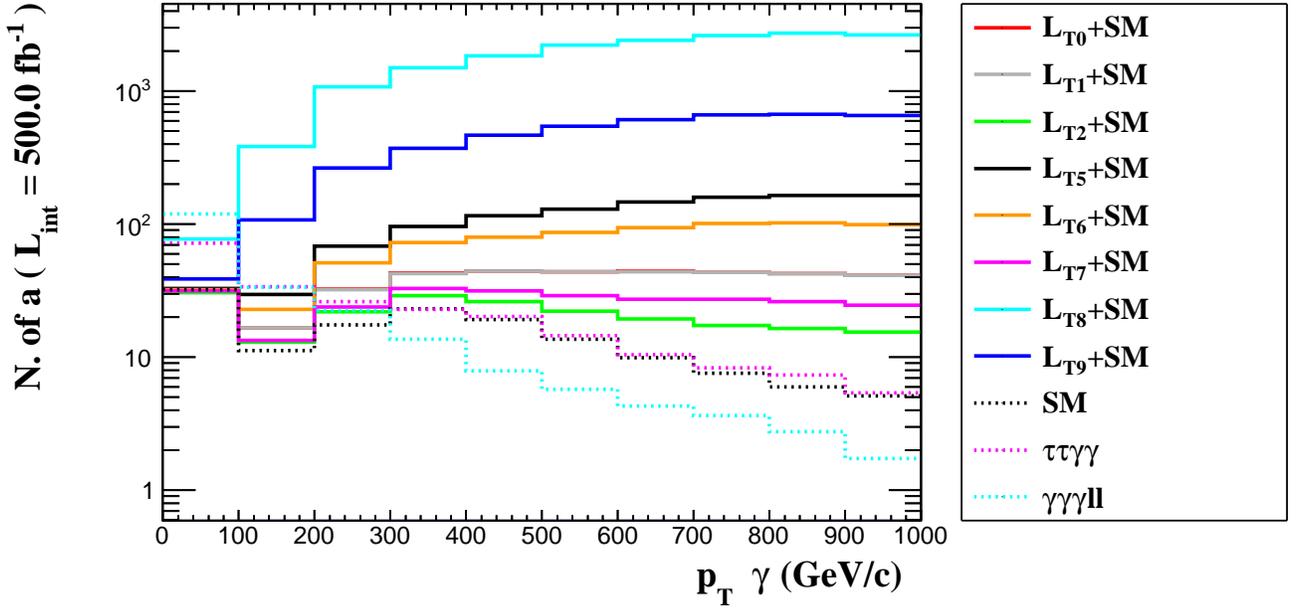}}}
\caption{ \label{fig:gamma}  The number of expected events as a function of $p^{\gamma}_T$
photon transverse momentum for the $e^+e^- \to Z\gamma\gamma$ signal and backgrounds at the
CLIC with $\sqrt{s}=3$ TeV and $P_{e^-}=0\%$. The distributions are
for $f_{T,j}/\Lambda^4$ with $j=0,1,2,5,6,7,8,9$ and relevant backgrounds. We considered
the contribution of SM, EFT and SM-EFT interference terms.}
\end{figure}

\begin{figure}[H]
\centerline{\scalebox{1.3}{\includegraphics{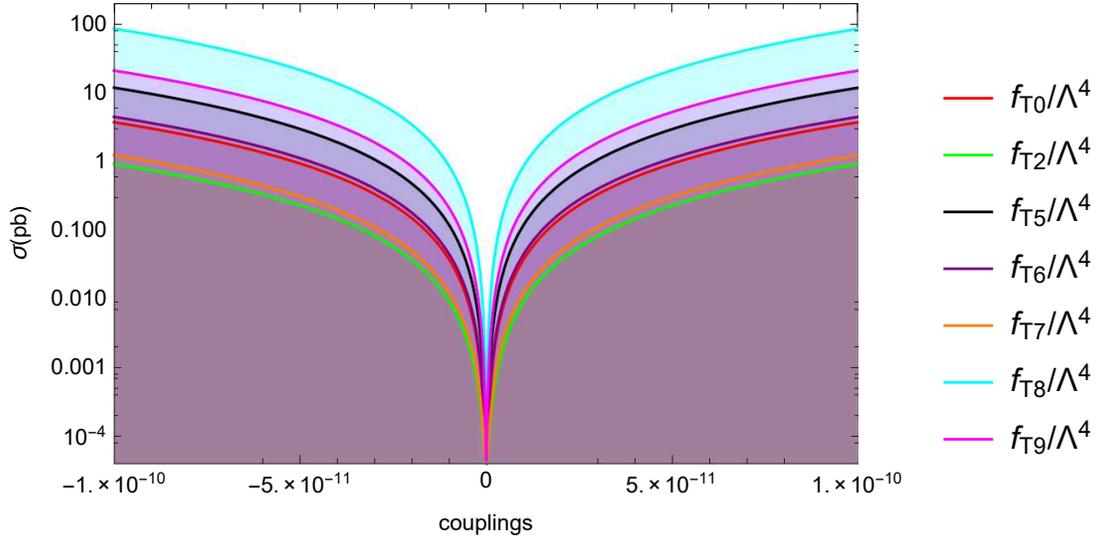}}}
\caption{ \label{fig:gamma} Production cross-section for the process $e^+e^-\to Z\gamma\gamma$
in terms of the aQGC $f_ {T,j}/\Lambda^4$ for the CLIC with $\sqrt{s}=3$ TeV and polarized beams $P_{e^-}=-80\%$. We considered the contribution of SM, EFT and SM-EFT interference
terms to the total cross-section.}
\end{figure}

\begin{figure}[H]
\centerline{\scalebox{1.25}{\includegraphics{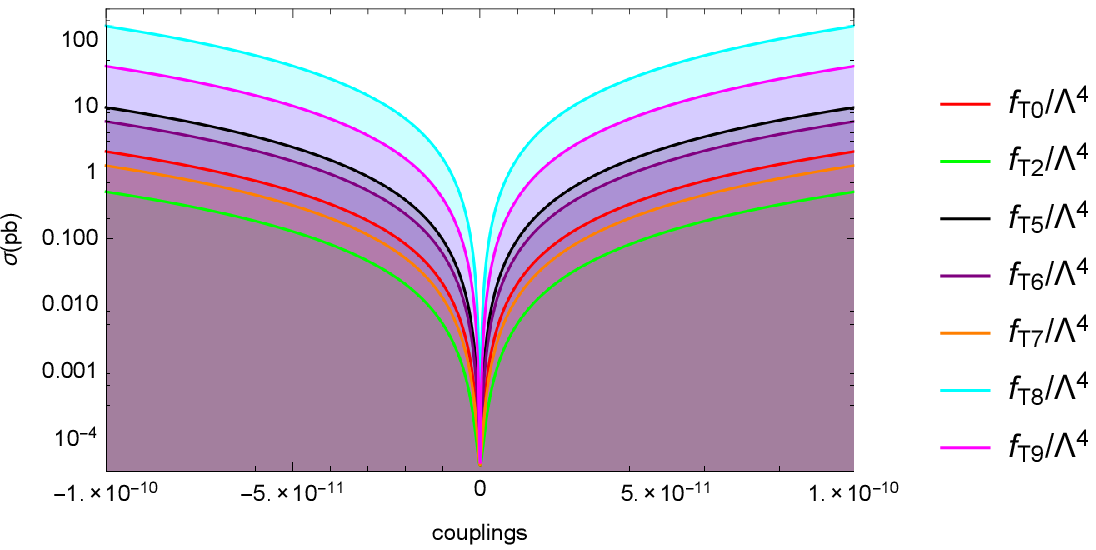}}}
\caption{ \label{fig:gamma} Same as in Fig. 3, but for $P_{e^-}=0\%$.}
\end{figure}

\begin{figure}[H]
\centerline{\scalebox{1.25}{\includegraphics{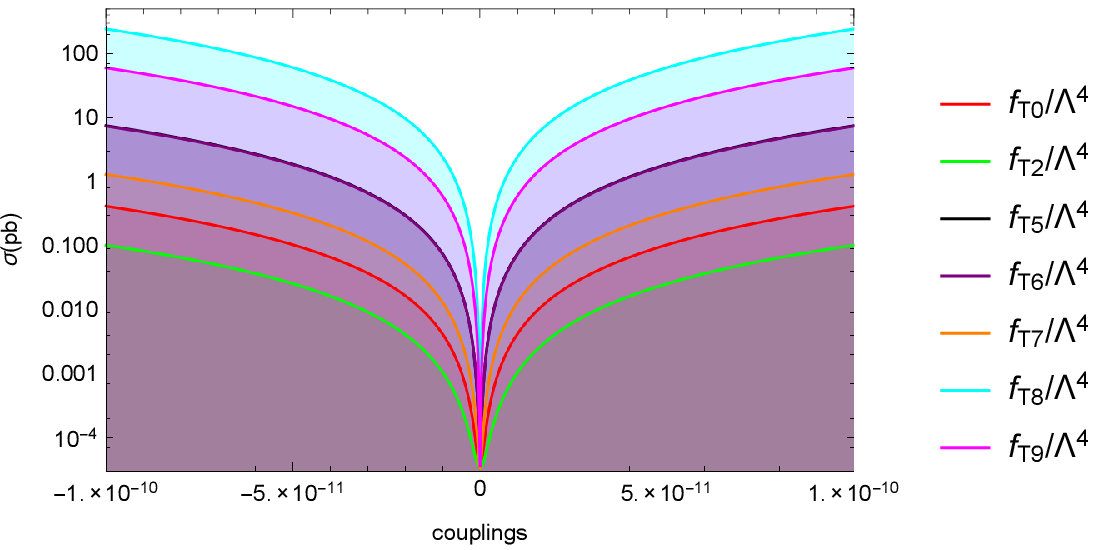}}}
\caption{ \label{fig:gamma} Same as in Fig. 3, but for $P_{e^-}=80\%$.}
\end{figure}

\begin{figure}[t]
\centerline{\scalebox{1.3}{\includegraphics{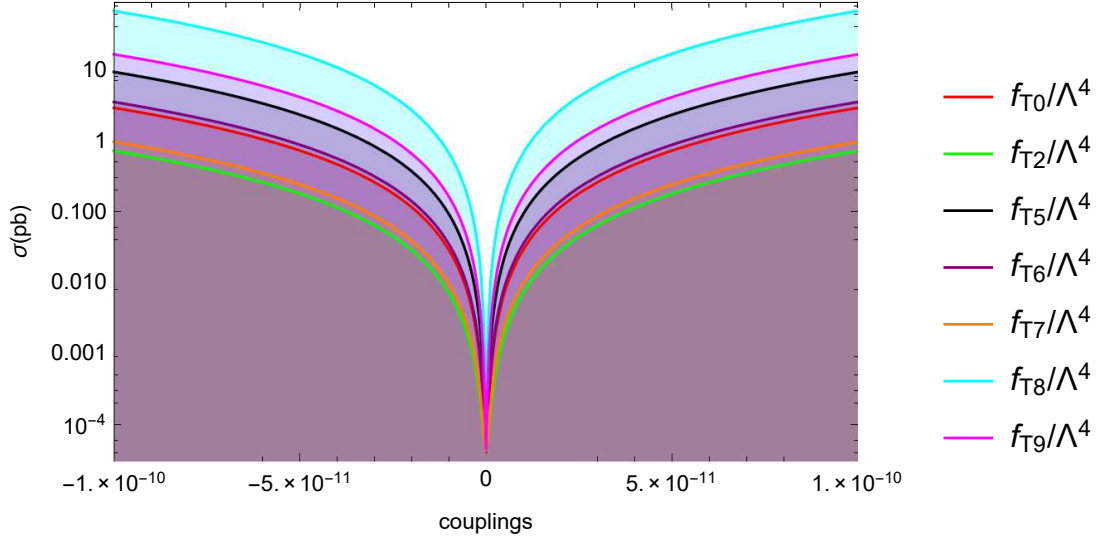}}}
\caption{ \label{fig:gamma} Production cross-section for the process $e^+e^-\to Z\gamma\gamma$ considering
ISR and Beamstrahlung effects in terms of the aQGC $f_ {T,j}/\Lambda^4$ for the CLIC with $\sqrt{s}=3$ TeV and polarized beams $P_{e^-}=-80\%$.}
\end{figure}

\begin{figure}[t]
\centerline{\scalebox{1.3}{\includegraphics{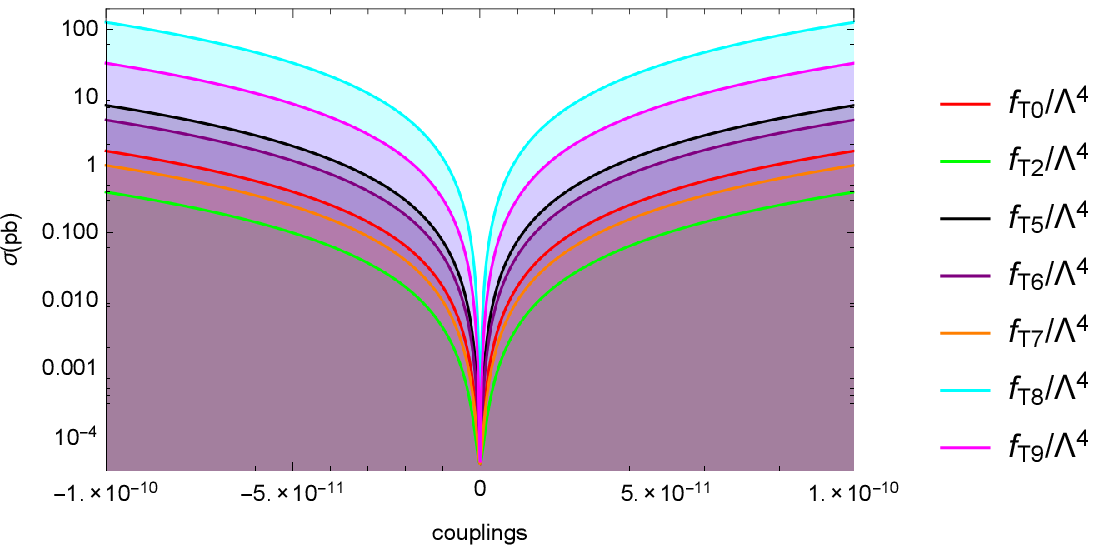}}}
\caption{ Same as in Fig. 6, but for $P_{e^-}=0\%$.}
\end{figure}

\begin{figure}[t]
\centerline{\scalebox{1.3}{\includegraphics{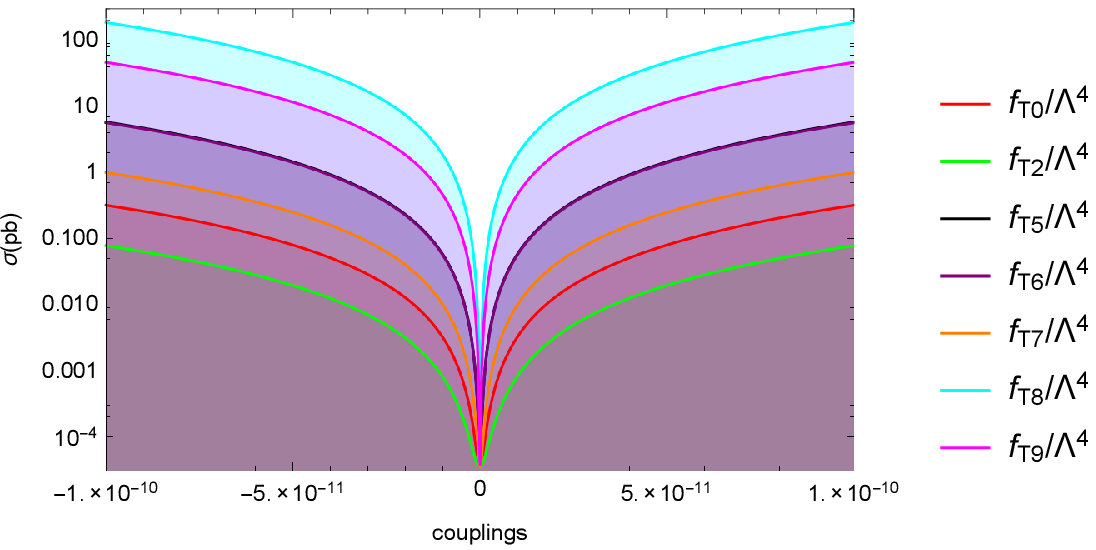}}}
\caption{ Same as in Fig. 6, but for $P_{e^-}=80\%$.}
\end{figure}

\begin{figure}[H]
\centerline{\scalebox{0.8}{\includegraphics{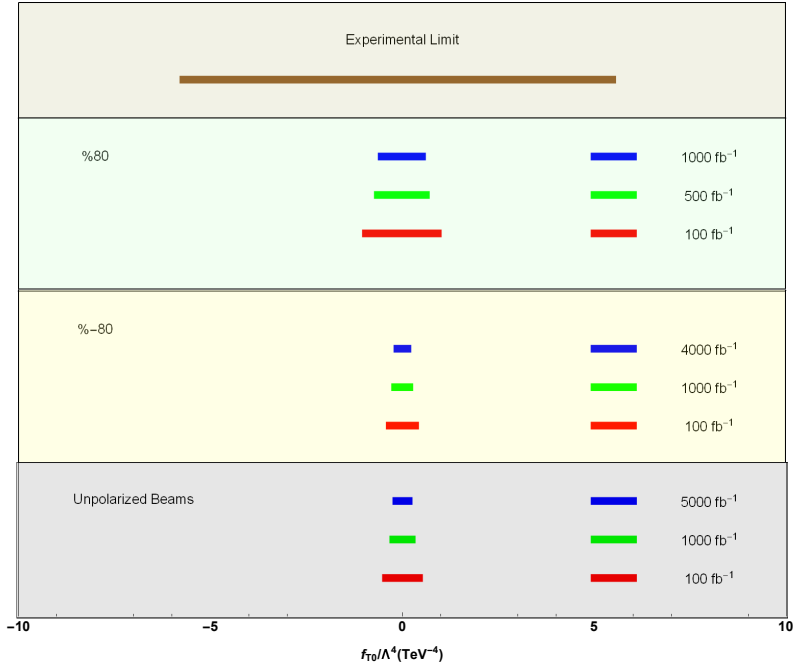}}}
\caption{ \label{fig:gamma} Comparison of the current experimental limits and projected sensitivity on $f_{T,0}/\Lambda^4$
for expected luminosities of ${\cal L}=100, 500, 1000, 4000, 5000\hspace{0.8mm} fb^{-1}$ and $\sqrt{s}=3\hspace{0.8mm}{\rm TeV}$
at the CLIC. We consider $P_{e^-}=-80\%, 0\%, 80\%$. We considered the contribution of SM, EFT and SM-EFT interference
terms.}
\end{figure}

\begin{figure}[H]
\centerline{\scalebox{0.8}{\includegraphics{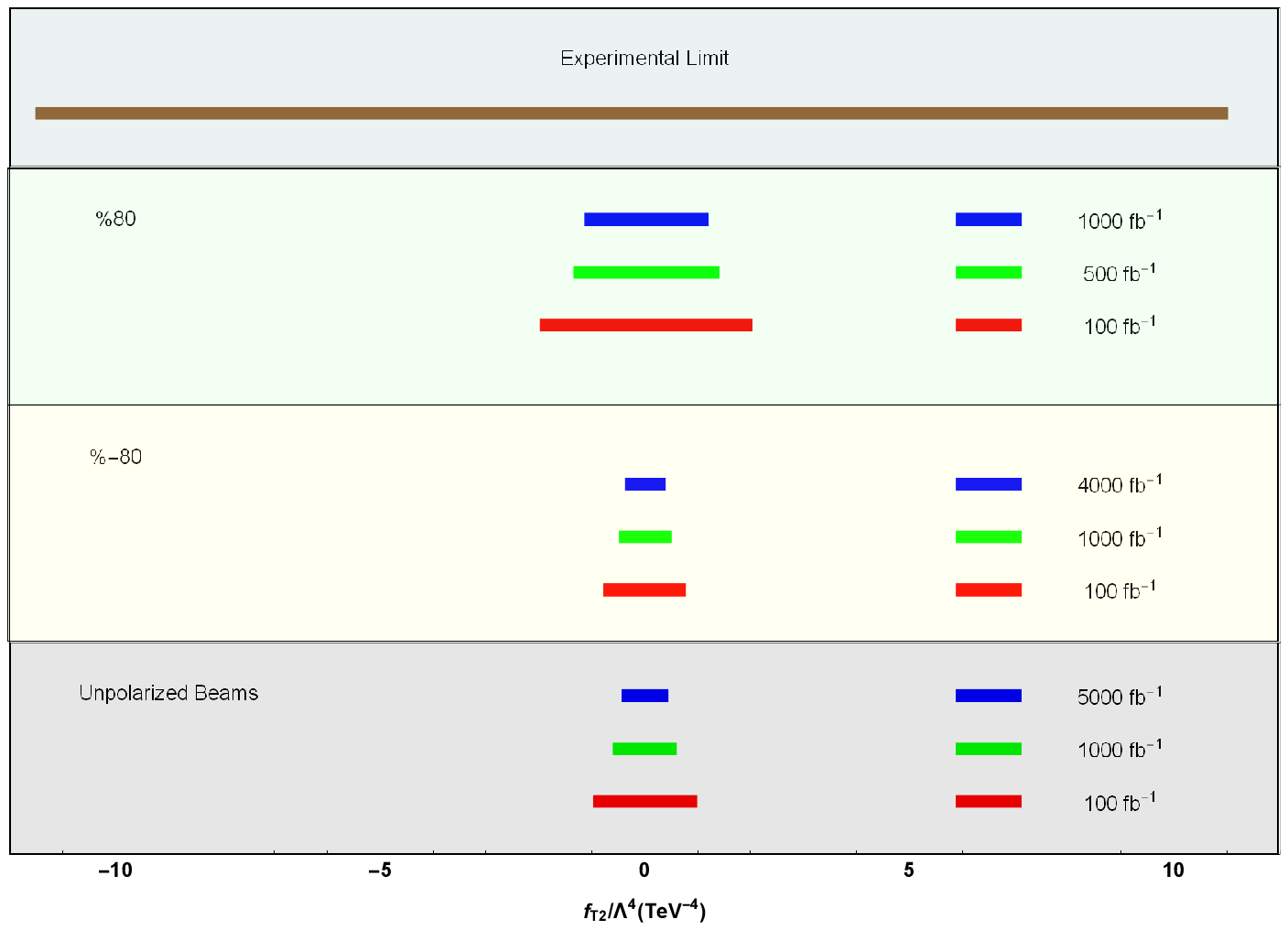}}}
\caption{ \label{fig:gamma} Same as in Fig. 9, but for $f_{T,2}/\Lambda^4$.}
\end{figure}

\begin{figure}[H]
\centerline{\scalebox{0.8}{\includegraphics{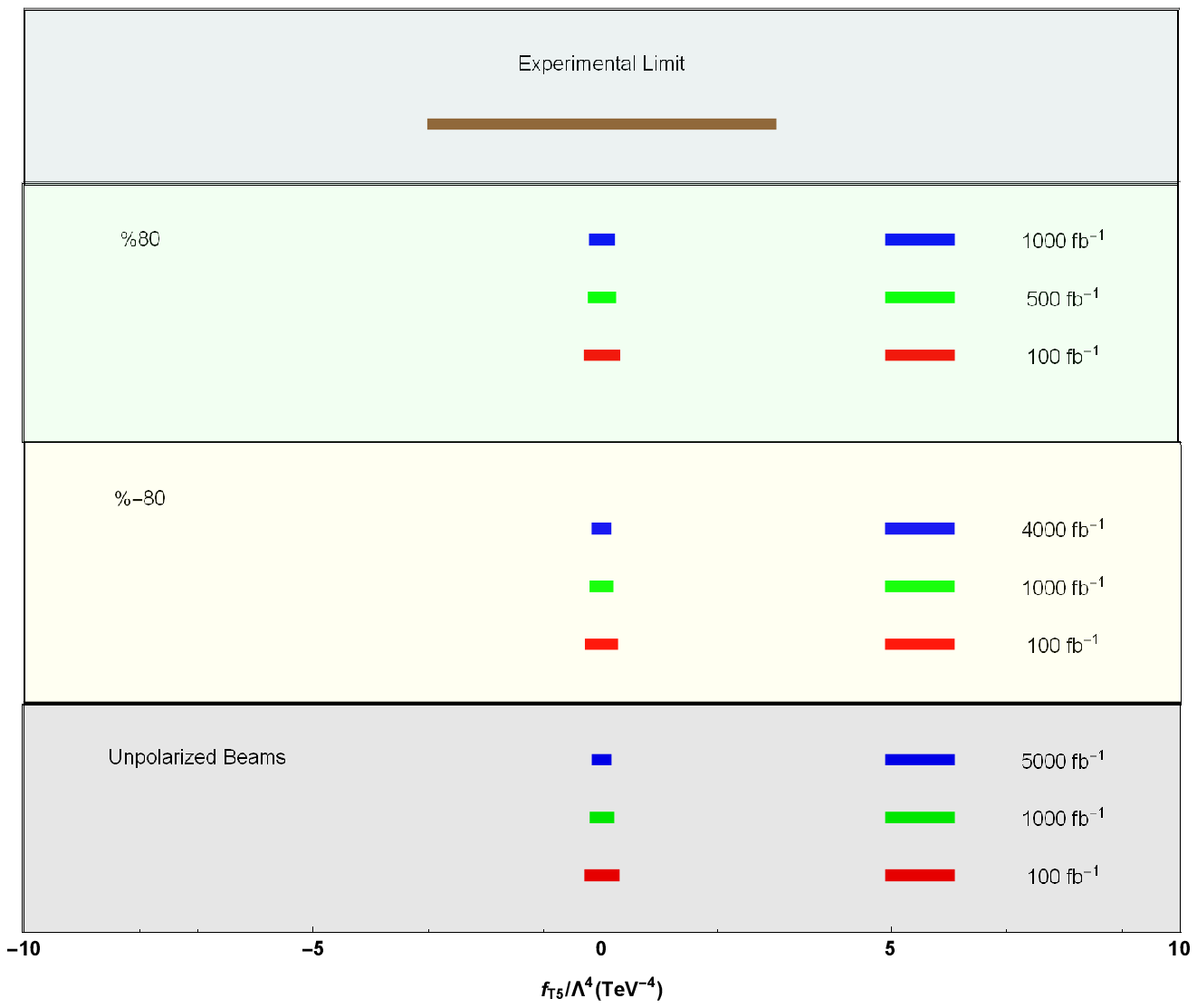}}}
\caption{ \label{fig:gamma}  Same as in Fig. 9, but for $f_{T,5}/\Lambda^4$.}
\end{figure}

\begin{figure}[H]
\centerline{\scalebox{0.73}{\includegraphics{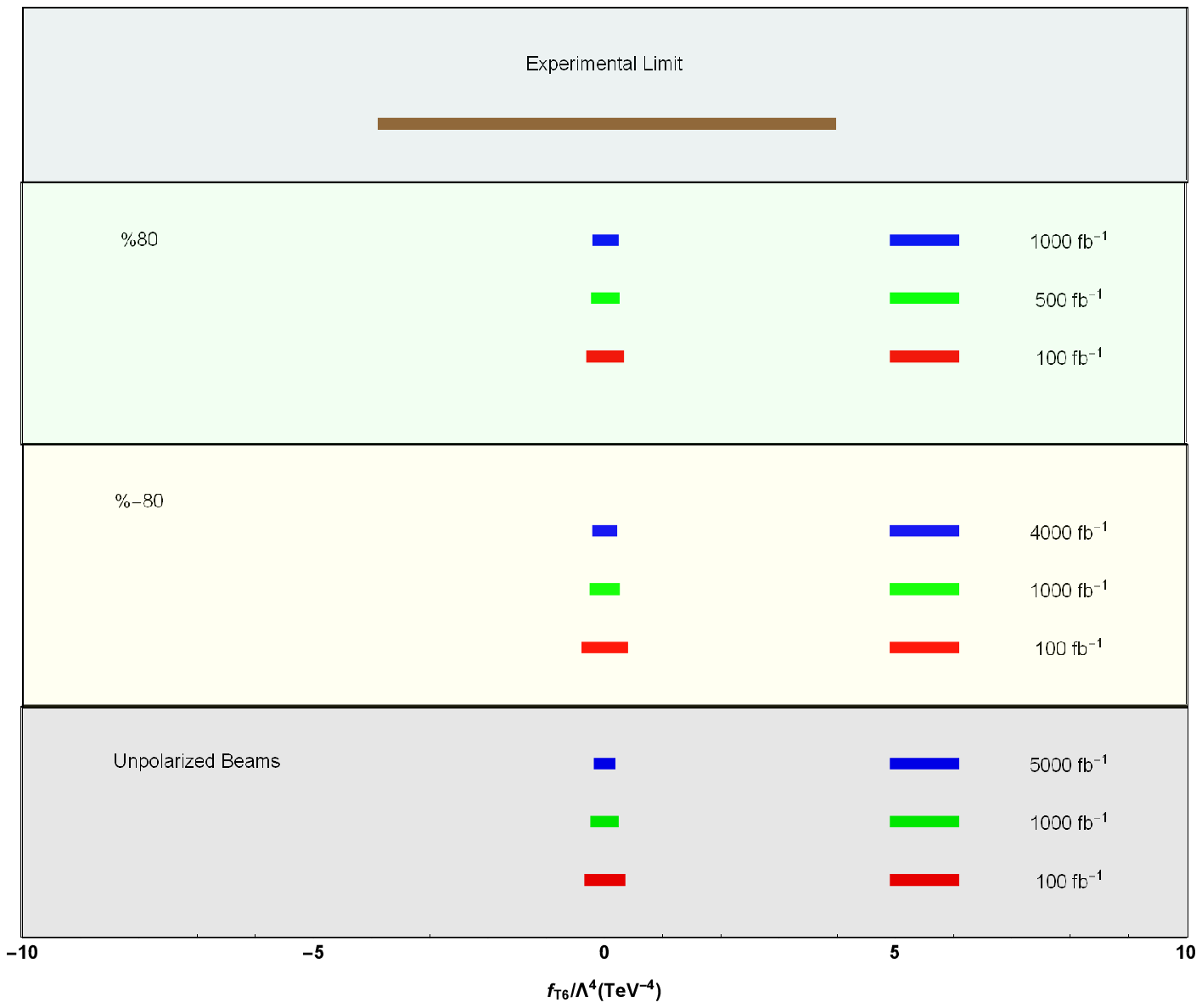}}}
\caption{ \label{fig:gamma} Same as in Fig. 9, but for $f_{T,6}/\Lambda^4$.}
\end{figure}

\begin{figure}[H]
\centerline{\scalebox{0.73}{\includegraphics{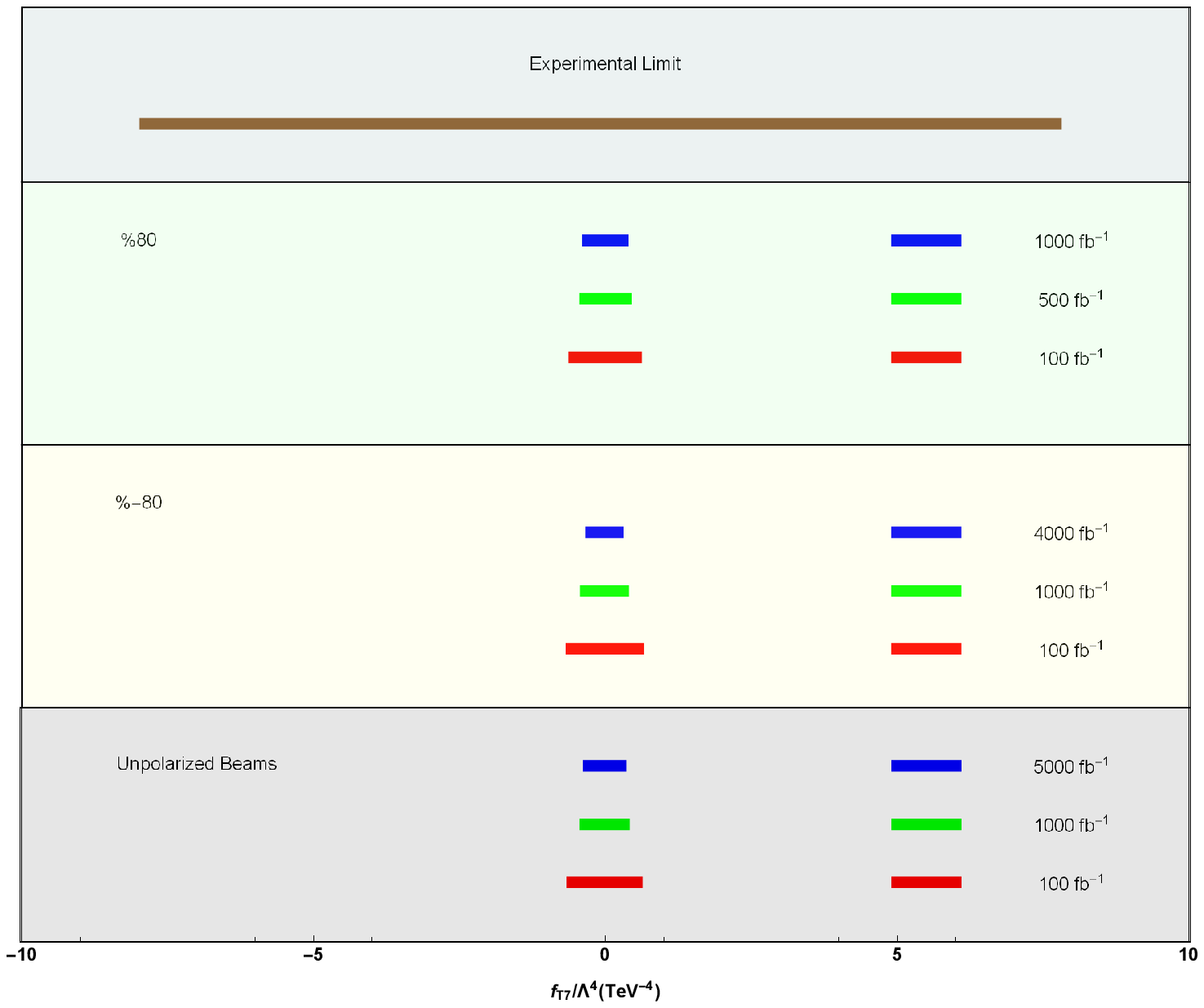}}}
\caption{ \label{fig:gamma} Same as in Fig. 9, but for $f_{T,7}/\Lambda^4$.}
\end{figure}

\begin{figure}[H]
\centerline{\scalebox{0.73}{\includegraphics{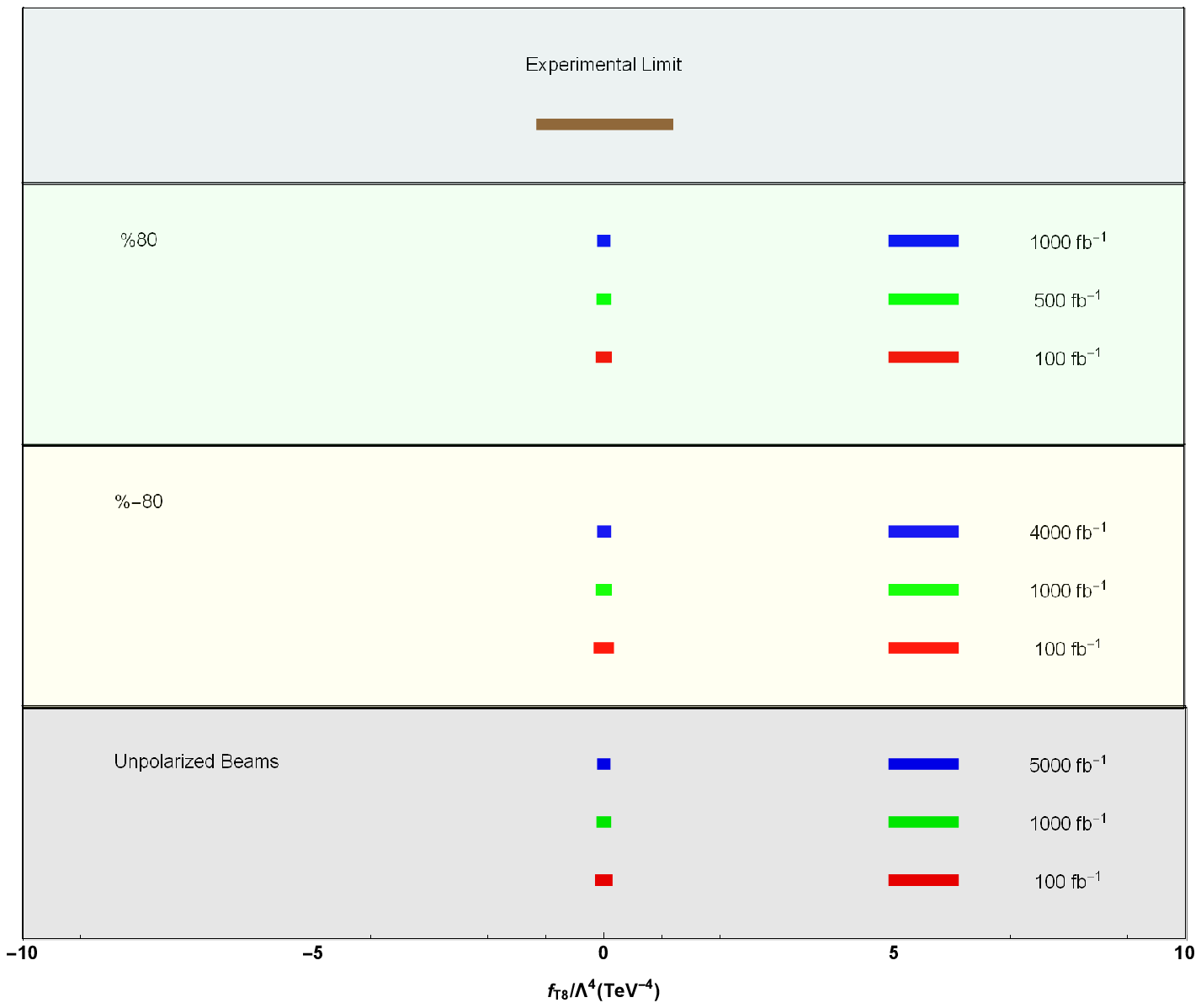}}}
\caption{ \label{fig:gamma} Same as in Fig. 9, but for $f_{T,8}/\Lambda^4$.}
\end{figure}

\begin{figure}[H]
\centerline{\scalebox{0.73}{\includegraphics{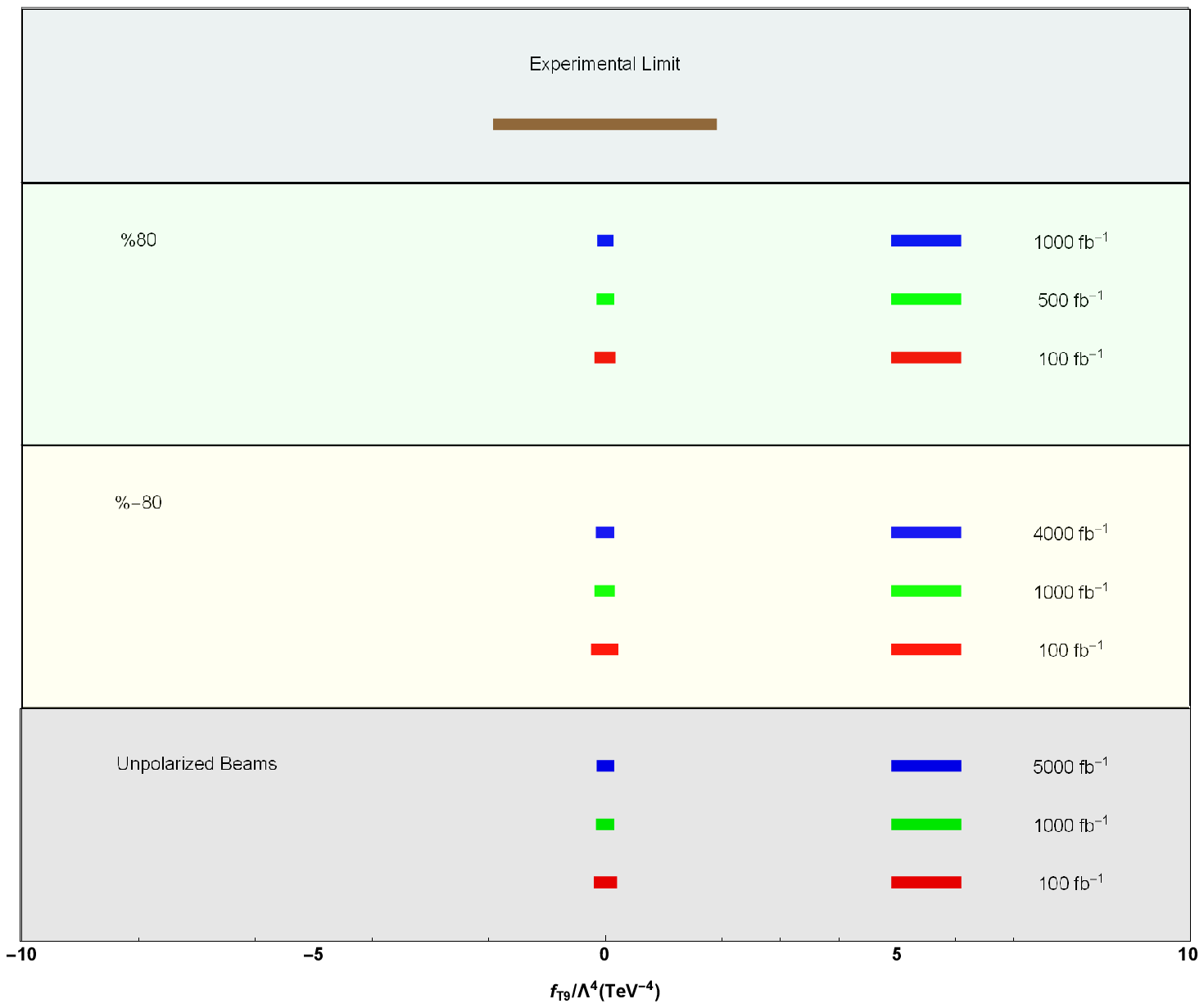}}}
\caption{ \label{fig:gamma} Same as in Fig. 9, but for $f_{T,9}/\Lambda^4$.}
\end{figure}

\end{document}